\RequirePackage{silence}
\RequirePackage[colorlinks=true, filecolor=black, citecolor=blue, linkcolor=blue, urlcolor=blue]{hyperref}
\documentclass[draft]{agujournal2019}
\usepackage{url} %this package should fix any errors with URLs in refs.
\usepackage{lineno}
\usepackage[inline]{trackchanges} % for better track changes.
\usepackage{soul}
\usepackage{multirow}           % Used in table
\usepackage{adjustbox}          % For floating
\usepackage{amsmath, amssymb}   % Mathematical expressions
\graphicspath{{./}{Figs/}}      % Figure pathœ
\newcommand{\orcid}[1]{\href{https://orcid.org/#1}{{\includegraphics[width=8pt]{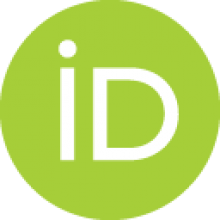}}}}
\usepackage[normalem]{ulem}

\newcommand{\textBoldadd}[1]{#1}
\newcommand{\textBoldmod}[2]{#2}
\newcommand{\textBoldrm}[1]{}

\journalname{Space Weather}

\begin{document}

%%%%%%%%%%%%%%%%%%%%%%%%%%%%%%%%%%%%%%%%%%%%%%%
%  TITLE
%%%%%%%%%%%%%%%%%%%%%%%%%%%%%%%%%%%%%%%%%%%%%%%
\title{Simulated Operational Testing of the Prototype Implementation of the SOFIE Model: The 2025 Space Weather Prediction Testbed Exercise}

%%%%%%%%%%%%%%%%%%%%%%%%%%%%%%%%%%%%%%%%%%%%%%%
%  AUTHORS AND AFFILIATIONS
%%%%%%%%%%%%%%%%%%%%%%%%%%%%%%%%%%%%%%%%%%%%%%%
\authors{
% Group 1: Modelers from UMich + SEPVAL [Contributions]
    Weihao Liu\affil{1}\orcid{0000-0002-2873-5688}, 
    Lulu Zhao\affil{1}\orcid{0000-0003-3936-5288}, 
    Igor V. Sokolov\affil{1}\orcid{0000-0002-6118-0469}, 
    Kathryn Whitman\affil{2,3}\orcid{0000-0002-3787-1622}, 
    Tamas I. Gombosi\affil{1}\orcid{0000-0001-9360-4951}, 
    Nishtha Sachdeva\affil{1}\orcid{0000-0001-9114-6133}, 
% Group 2: SWPC & CCMC & SRAG & M2M hosts [Last Name Initials]
    Eric T. Adamson\affil{4}\orcid{0000-0002-2125-0346}, 
    Hazel M. Bain\affil{4,5}\orcid{0000-0003-2595-3185}, 
    Claudio Corti\affil{6,7,8}\orcid{0000-0001-9127-7133}, 
    M. Leila Mays\affil{6}\orcid{0000-0001-9177-8405}, 
    Michelangelo Romano\affil{6,9}\orcid{0000-0001-6258-752X}, 
% Group 3: Participants of the testbed exercise [Last Name Initials]
    Carina R. Alden\affil{6,9}\orcid{0000-0001-7330-8792}, 
    Madeleine M. Anastopulos\affil{6,9}\orcid{0000-0001-5123-7188}, 
    Mary E. Aronne\affil{6,10}, 
    Janet E. Barzilla\affil{2,11}\orcid{0000-0002-2522-1024}, 
    Wesley T. Cook\affil{4}, 
    Shawn D. Dahl\affil{4}, 
    % Ricky A. Egeland\affil{2}\orcid{0000-0002-4996-0753}, 
    Hannah Hermann\affil{6,9}, 
    Anthony J. Iampietro\affil{6,9}, 
    A. Steve Johnson\affil{2,11}, 
    Elizabeth A. Juelfs\affil{6,10}, 
    Melissa R. Kane\affil{6,9}\orcid{0009-0005-7792-2585}, 
    Jonathan D. Lash\affil{4}, 
    Kimberly Moreland\affil{4,5}\orcid{0000-0002-6202-8565}, 
    Briana K. Muhlestein\affil{4}, 
    Teresa Nieves-Chinchilla\affil{6}\orcid{0000-0003-0565-4890}, 
    % Alessandra A. Pacini\affil{4}\orcid{0000-0003-0187-1652}, 
    Edward Semones\affil{2}, 
    James F. Spann\affil{12}\orcid{0000-0002-0176-3259}, 
    Earl M. Spencer\affil{4}, 
    Luke A. Stegeman\affil{2,11}\orcid{0000-0003-0716-4465}, 
    Christopher J. Stubenrauch\affil{6,9}\orcid{0000-0002-0566-5912}, 
    Kenneth L. Tegnell\affil{4} 
}

\affiliation{1}{Department of Climate and Space Sciences and Engineering, University of Michigan, Ann Arbor, MI 48109, USA}
\affiliation{2}{Space Radiation Analysis Group, NASA Johnson Space Center, Houston, TX 77058, USA}
\affiliation{3}{KBR, Houston, TX 77002, USA}
\affiliation{4}{NOAA Space Weather Prediction Center, Boulder, CO 80305, USA}
\affiliation{5}{Cooperative Institute for Research In Environmental Sciences, University of Colorado, Boulder, CO 80309, USA}
\affiliation{6}{Heliophysics Science Division, NASA Goddard Space Flight Center, Greenbelt, MD 20771, USA}
\affiliation{7}{Universities Space Research Association, Washington, DC 20024, USA}
\affiliation{8}{Physics and Astronomy Department, University of Hawaii at Manoa, Honolulu, HI 96822, USA}
\affiliation{9}{Department of Physics, The Catholic University of America, Washington, DC 20064, USA}
\affiliation{10}{Department of Physics and Astronomy, George Mason University, Fairfax, VA 22030, USA}
\affiliation{11}{Leidos, Inc., Webster, TX 77598, USA}
\affiliation{12}{NOAA NESDIS Office of Space Weather Observations, Silver Spring, MD 20706, USA}

% Corresponding author mailing address and e-mail address:
\correspondingauthor{Weihao Liu}{whliu@umich.edu}

%%%%%%%%%%%%%%%%%%%%%%%%%%%%%%%%%%%%%%%%%%%%%%%
% KEY POINTS
%%%%%%%%%%%%%%%%%%%%%%%%%%%%%%%%%%%%%%%%%%%%%%%
%  List up to three key points (at least one is required)
%  Key Points summarize the main points and conclusions of the article
%  Each must be 140 characters or fewer with no special characters or punctuation and must be complete sentences

\begin{keypoints} % 140 | 134 | 136 now
    \item A prototype of SOFIE, an integrated physics-based model, was run on-site in the 2025 SWPT exercise under simulated operational conditions.
    % \item SOFIE \textBoldmod{predicted solar energetic particle}{predicts SEP} fluxes \textBoldadd{accurately and} significantly faster than real time\textBoldmod{ while reproducing key features of observed fluxes}{, although the earliest SEP forecast requires a few hours}.
    \item SOFIE predicts SEP fluxes significantly faster than real time, although the earliest SEP forecast requires a few hours after CME onset.
    \item Model optimization and forecaster's feedback pave the way for SOFIE's operational utility, supporting future space exploration missions.
\end{keypoints}

%%%%%%%%%%%%%%%%%%%%%%%%%%%%%%%%%%%%%%%%%%%%%%%
%  ABSTRACT and PLAIN LANGUAGE SUMMARY
% see http://sharingscience.agu.org/creating-plain-language-summary/)
%%%%%%%%%%%%%%%%%%%%%%%%%%%%%%%%%%%%%%%%%%%%%%%

\begin{abstract}
The CLEAR Space Weather Center of Excellence's solar energetic particle (SEP) model, SOlar wind with FIeld lines and Energetic particles (SOFIE), was run and evaluated on-site during the Space Weather Prediction Testbed (SWPT) exercise at the National Oceanic and Atmospheric Administration's Space Weather Prediction Center (NOAA/SWPC) in May 2025. 
As a physics-based SEP model, SOFIE simulates the acceleration and transport of energetic particles by the coronal mass ejection (CME)-driven shock in the solar corona and inner heliosphere, and has been validated against historical events. However, questions remain regarding whether a physics-based model, traditionally considered computationally expensive, could meet operational needs. 
The SWPT exercise offered a valuable opportunity to evaluate SOFIE under simulated operational conditions. On-site interactive feedback from SWPC forecasters, Space Radiation Analysis Group (SRAG) console operators, Community Coordinated Modeling Center (CCMC) personnel, and Moon-to-Mars Space Weather Analysis Office (M2M SWAO) analysts led to significant strategic improvements in the model configuration. The simulation grid was optimized by combining a coarser background grid with higher-resolution regions along the CME path and toward Earth, reducing computational cost without compromising accuracy. 
In this work, we present the simulated operational performance of SOFIE and its capability to predict SEP fluxes significantly faster than real time. During the SWPT exercise, SOFIE completed a 4-day SEP simulation within 5 hours using 1,000 central processing unit cores, although the earliest SEP forecast was obtained a few hours after CME onset. This marks a milestone in demonstrating SOFIE's operational usefulness and robustness to support future human space exploration.
\end{abstract}

\section*{Plain Language Summary}
% Enter your Plain Language Summary here or delete this section.
% Here are instructions on writing a Plain Language Summary: 
% https://www.agu.org/Share-and-Advocate/Share/Community/Plain-language-summary

Solar energetic particles (SEPs) are bursts of high-energy particles released by the Sun, often during large eruptions called coronal mass ejections (CMEs). These particles can reach Earth in less than an hour, posing hazards to satellites, astronauts, and even ground-based technologies. Therefore, reliable SEP predictions are critical for future human space missions. 
In this work, we describe the operational testing of the SOlar wind with FIeld lines and Energetic particles (SOFIE) model during the Space Weather Prediction Testbed (SWPT) exercise at the National Oceanic and Atmospheric Administration's Space Weather Prediction Center (NOAA/SWPC). SOFIE is an integrated physics-based model that simulates the solar wind, CME generation and propagation, and shock-driven acceleration and transport processes of SEPs. 
During the SWPT exercise, SOFIE was run on 1,000 central processing unit cores for two historical events under simulated operational conditions. It reproduced key features of CME coronagraph images and SEP fluxes, delivering 4-day predictions significantly faster than real time, although the earliest SEP forecast was produced a few hours after CME onset. This demonstrates that SOFIE as a physics-based SEP model, although traditionally considered computationally expensive, can provide accurate and real-time SEP predictions, marking an important step toward operational usefulness for future space exploration.

%%%%%%%%%%%%%%%%%%%%%%%%%%%%%%%%%%%%%%%%%%%%%%%
%  BODY TEXT
%%%%%%%%%%%%%%%%%%%%%%%%%%%%%%%%%%%%%%%%%%%%%%%
\section{Motivation} \label{sec0:motive}

% Para 1: SWPT Exercise at NOAA/SWPC - What is it
In April and May 2025, the National Oceanic and Atmospheric Administration's Space Weather Prediction Center (NOAA/SWPC), together with the Community Coordinated Modeling Center (CCMC), the National Aeronautics and Space Administration's Space Radiation Analysis Group (NASA/SRAG), and the NASA Moon-to-Mars Space Weather Analysis Office (M2M SWAO), hosted the first Space Weather Prediction Testbed (SWPT) exercise (\url{https://testbed.spaceweather.gov/exercises/2025-artemis-ii-human-spaceflight-support-exercise-information}) in support of the Artemis II mission, marking a major leap forward in operational readiness for human spaceflight. 
The exercise brought together SWPC forecasters, SRAG console operators, M2M SWAO analysts, CCMC personnel, and experts from commercial industry and academia to evaluate space weather products and operational workflows. %The outcome exceeded expectations across technical performance, organizational coordination, and strategic value, demonstrating a modern, agile capability to deliver timely, mission-critical space weather information to safeguard astronauts and spacecraft. 

% Para 2: What we (all) did in the SWPT exercise
Within this exercise, SWPC, M2M SWAO, CCMC, and SRAG played distinct, but complementary roles. SWPC serves as the official civilian operational forecast center and their forecasters provided the initial solar event alerts, including flare detections. Both SWPC and M2M SWAO characterized the associated coronal mass ejections (CMEs) through the SWPC CME Analysis Tool (SWPC-CAT) and provided these analyses (though the M2M SWAO-derived values were used for our subsequent modeling of the eruptive event). CCMC provides research-to-operations model testing and validation support, and SRAG console operators assess radiation exposure in real time. 
%This collaboration exemplifies how expertise from each organization can be leveraged to enhance space weather preparedness for future missions. 
Specifically, participants worked in small teams through two solar radiation storm scenarios based on the 10 September 2017 and 4 November 2001 solar energetic particle (SEP) events, two strong, well-observed historical events observed at Earth, accompanied by contextual observations of the solar wind, flares, CMEs and SEPs. %During a human space exploration mission like Artemis II, SEP events such as these would require careful monitoring by SRAG operators. 
Radiation levels for both events were simulated inside the Artemis II Orion vehicle: while doses observed inside the vehicle over the course of the simulated events would not pose acute health risks to astronauts, they would exceed the threshold to build shelter, giving SRAG console operators the opportunity to describe the planned space radiation response to Testbed observers. For each event, the exercise examined organizational coordination, technical performance, and strategic decision making, and also tested the accuracy and runtime performance of SEP models for real-time SEP predictions. 

% Para 3: What we (CLEAR team) did and this Paper Structure
The NASA's Space Weather Center of Excellence, CLEAR, has actively participated in the May 2025 SWPT exercise. During the exercise, the CLEAR team conducted on-site simulations using the SOlar wind with FIeld lines and Energetic particles \cite<SOFIE,>[]{zhao2024solar, liu2025physics} model to test its ability to predict SEP fluxes in real time. % In-depth discussions with personnel from SWPC, SRAG, CCMC and M2M office led to strategic improvements of SOFIE's setup for operations, providing valuable feedback on its predictive performance and enhancing the model's readiness for operational applications. 
In the following, we present the results of SOFIE's application during the SWPT exercise. In Section~\ref{sec1:intro}, we provide a background introduction of SEPs. Section~\ref{sec2:method} describes the workflow of the SOFIE model. Sections~\ref{sec3:2017} and \ref{sec4:2001} present the two representative historical SEP events selected for this exercise, the 10 September 2017 and 4 November 2001 eruptions, comparing model results with spacecraft observations. In Section~\ref{sec5:discuss}, we discuss the operational implications of SOFIE, including lessons learned from the SWPT exercise and the balance between grid resolution and computational efficiency. Finally, Section~\ref{sec6:sumcon} summarizes the findings and outlines future directions toward operational SEP prediction capabilities.

\section{Radiation Hazard from SEPs} \label{sec1:intro}

% Para 1: SEP - Mechanisms - Radiation Risks - Important to Space Weather
SEPs are one of the most hazardous manifestations of solar activity for both space- and ground-based systems. 
Accelerated by shock-driven acceleration and/or magnetic reconnection processes in flares and coronal jets \cite{desai2016large, klein2017acc}, these high-energy particles are observed to span a wide range of energies from tens of keV to a few GeV, and their spectra evolve dynamically during an event. 
They can penetrate deep into spacecraft shielding or human tissue, causing short-term increases in radiation levels that can last up to several days \cite<e.g.,>[]{mertens2019characterization, guo2021radiation, buzulukova2022space, krittanawong2022human}. 
For human spaceflight beyond Earth's magnetosphere, SEPs can also threaten satellite electronics, communications, and pose significant radiation risks with radiation levels that can rise rapidly within minutes to hours \cite<e.g.,>[]{minow2020space, lowe2025nowcasting}. Consequently, SEP predictions have long been recognized as a crucial element of space weather operations \textBoldadd{\cite{anastasiadis2019solar, georgoulis2024prediction, guo2024particle, papaioannou2025predicting}}. 
% Despite decades of efforts, accurately predicting the timing and intensity of SEP events remains a formidable challenge, requiring both an understanding of underlying physical processes and tools that can deliver timely forecasts under operational constraints \cite{guo2024particle, zheng2024overview}. 

% Para 2: SEP - Operation Interests and Energies
In operational practice, NOAA/SWPC radiation storm services are traditionally based on proton intensities in geostationary orbit, as measured by particle detectors on board the Geostationary Operational Environmental Satellites \cite<GOES,>[]{menzel1994introducing, onsager1996operational, rodriguez2014intercalibration, kress2020goes}. The corresponding SWPC radiation products are based on the NOAA solar radiation storm scale (\url{https://www.swpc.noaa.gov/noaa-scales-explanation}), which relates the GOES $>$10 MeV proton flux to impacts on humans, satellite systems, navigation, and other technologies \cite{bain2023noaa}. 
A solar particle event, defined as the $>$10 MeV proton flux exceeding 10 particle flux units (pfu, $\#\; \mathrm{cm^{-2}\; s^{-1}\; sr^{-1}}$), poses a potential radiation hazard to astronauts during extravehicular activity and yields an increased probability of single-event upsets in satellites. In addition, the $>$100 MeV proton flux exceeding 1 pfu, suggesting an energetic solar particle event, is used as another threshold, marking conditions associated with increased astronaut radiation exposure even inside spacecraft shielding. In the following SEP flux comparisons, we focus on these two operational energy channels when evaluating SOFIE's predictive performance against GOES observations.

\section{The SOFIE Model} \label{sec2:method}

\begin{figure*}[ht!]
    \centering{
    \includegraphics[width=0.9\textwidth]{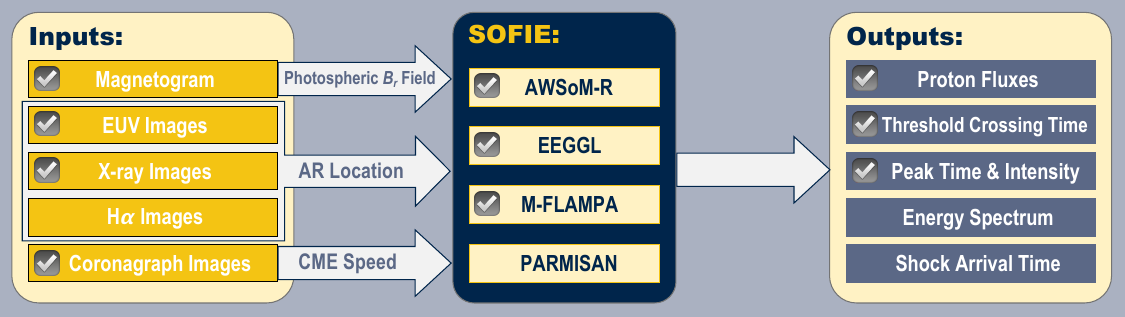}}
    \caption{Schematic diagram of the prototyped SOFIE model suite (middle), as well as its inputs (left) and SEP outputs (right). Items with a marker were used in the SWPT exercise.} \label{fig:sofie}
    \end{figure*}

SOFIE is an integrated solar corona--interplanetary medium--CME--SEP model suite, with the schematic diagram in Figure \ref{fig:sofie} describing its workflow. Built within the Space Weather Modeling Framework \cite<SWMF,>[]{toth2005space, toth2012adaptive, gombosi2021sustained}, SOFIE integrates four main components: 
(1) the Alfv\'{e}n Wave Solar atmosphere Model-Realtime \cite<AWSoM-R,>[]{van2014alfven, sokolov2013magnetohydrodynamic, sokolov2021threaded}, solving stream-aligned magnetohydrodynamic (MHD) equations \cite{sokolov2022stream}, designed for simulating the ambient solar wind with reliable magnetic connectivity; 
(2) the Eruptive Event Generator with the Gibson-Low magnetic configuration \cite<EEGGL,>[]{gibson1998time, borovikov2017eruptive, jin2017data} for CME generation and propagation; 
(3) the Multiple Field-Line-Advection Model for Particle Acceleration \cite<M-FLAMPA,>[]{sokolov2004new, borovikov2018toward, liu2025physics}, used to simulate particle acceleration and transport processes; 
and (4) the Particle ARizona and MIchigan Solver on Advected Nodes \cite<PARMISAN,>[]{chen2025evidence}, a Monte-Carlo solver for SEPs. 
\textBoldadd{While PARMISAN is a test-particle code that needs more than millions of test particles for good statistics, M-FLAMPA directly solves the canonical distribution function of SEPs from the Parker transport equation \cite{parker1965passage}. In M-FLAMPA, the advecting effect of solar wind  flows, adiabatic cooling/heating, and parallel diffusion are incorporated, while the cross-field diffusion and particle drift (gradient, curvature, and current sheet drift) effects are neglected, with particle transport calculated along individual magnetic field lines only. Although this simplification limits the treatment of lateral transport across field lines, it allows M-FLAMPA to efficiently simulate particle acceleration at CME-driven shocks and the subsequent transport process, which is suitable for operational predictions.} 
\textBoldmod{We}{Hence, we} used the first three components of SOFIE in the testbed exercise\textBoldadd{, as presented in Figure \ref{fig:sofie}}. \textBoldadd{A comprehensive description of the SOFIE model is available in \citeA{zhao2024solar} and \citeA{liu2025physics}.} %In the following, we introduce each module of SOFIE and the simulation setup used during the testbed exercise.

To simulate an SEP event, a magnetogram, which provides the photospheric radial magnetic field ($B_r$), is taken as input for AWSoM-R to simulate the three-dimensional (3D) global solar wind background. In AWSoM-R, the Poynting flux parameter serves as the primary free parameter for tuning the background solar wind, which depends on the phase of the solar cycle and is adjusted in operations based on Equation (1) in \citeA{huang2024solar}. 
When a CME eruption is observed, the extreme-ultraviolet (EUV), X-ray, or H$\alpha$ images are typically used to provide the active region (AR) location, with EUV and X-ray images adopted for this purpose in the testbed exercise. We also used coronagraph images to estimate the CME speed. The EEGGL module then combines the magnetogram, AR location, and CME speed to generate the CME flux rope, which is subsequently propagated in AWSoM-R MHD simulations, with M-FLAMPA modeling the resulting SEP fluxes simultaneously. 
In M-FLAMPA, there are mainly two free parameters: (1) the seed particle scaling factor, which sets the injected particle population at the shock front with a default value of 1.0 and can be adjusted to better reproduce observed profiles\textBoldadd{, when the SEP peak intensity is observed}; and (2) the upstream mean free path parameter \cite<see Equation (15) in>[]{liu2025physics}, which affects particle acceleration, transport and the resulting time--intensity profiles, and was set as 0.1 \textBoldmod{au}{astronomical units (au)} for both events during the SWPT exercise. 
Detailed discussions of their impacts and uncertainties are provided in Section \ref{sec5.2:uncertainty}.

From these coupled simulations, SOFIE provides the two-dimensional (2D) proton flux distribution on a Sun-centered sphere at a specified radial distance, threshold crossing time of operational channels, peak time and intensity, energy spectrum and shock arrival time, etc. In the testbed exercise, we show the 2D distribution on the 1 au sphere and extract the time--intensity profile along Earth's trajectory. 
In Table \ref{tab:param}, we list the SOFIE input parameters used in the testbed exercise and further discuss them in Sections \ref{sec3:2017} and \ref{sec4:2001} below. 

    \begin{table*}[ht!]
    \begin{center}
    \caption{Input parameters of the AWSoM-R, EEGGL, and M-FLAMPA models in SOFIE for the SWPT exercise simulation.} \label{tab:param}
    \setlength\tabcolsep{6pt}{
    \adjustbox{width=\textwidth}{
    \begin{tabular}{clccc}
        \hline\hline
        \multirow{2}{*}{Model} & \multirow{2}{*}{Input Parameter} & \multirow{2}{*}{Provider} & \multicolumn{2}{c}{Value for the Eruptive Event on} \\
        \cline{4-5}
        & & & 10 September 2017 & 4 November 2001 \\ 
        \hline % \cline{2-3} \\ \multirow{LINE}{*}{CONTENT} \multicolumn{NUMBER}{c}{xxx}
        AWSoM-R & Poynting flux parameter & Modeler & $0.55\; \mathrm{MW\; m^{-2}\; T^{-1}}$ & $0.20\; \mathrm{MW\; m^{-2}\; T^{-1}}$ \\
        \hline % \cline{2-3}
        EEGGL & AR location$^{a}$ & SWPC & S12W85 & N05W24 \\
        & CME speed & M2M SWAO & $2650\; \mathrm{km\; s^{-1}}$ & $1925\; \mathrm{km\; s^{-1}}$ \\
         % & Type of the inserted flux rope & Model & Spheromak & Spheromak \\
         % & AR center location$^{a}$ & & ($115.50^{\circ}, -11.24^{\circ}$) & ($135.45^{\circ}, 5.38^{\circ}$) \\
         % & Flux rope radius$^{b}$     & EEGGL & $0.30\; R_\mathrm{s}$ & $0.70\; R_\mathrm{s}$ \\
         % & Flux rope stretching$^{b}$ & EEGGL & $0.60\; R_\mathrm{s}$ & $0.56\; R_\mathrm{s}$ \\
         % & Flux rope height$^{b}$     & EEGGL & $0.50\; R_\mathrm{s}$ & $0.88\;R_\mathrm{s}$ \\
         % & Flux rope magnetic field strength & & $99.2\; \mathrm{G}$ & $40.0\; \mathrm{G}$ \\
        \hline
        M-FLAMPA & Injection scaling factor & Modeler & $1.0$ & $10.0$ \\ 
        & Mean free path parameter$^{b}$ & Modeler & $0.1\;\mathrm{au}$ & $0.1\;\mathrm{au}$ \\ 
        \hline
        \multicolumn{4}{l}{$^a$ These locations are given in Stonyhurst heliographic longitude and latitude.} \\
        % \multicolumn{4}{l}{$^b$ The unit of $R_\mathrm{s}$ refers to the solar radius.} \\
        \multicolumn{4}{l}{$^b$ This corresponds to $\lambda_0$ of Equation (15) in \citeA{liu2025physics}.} \\
    \end{tabular}}}
    \end{center}
    \end{table*}

\section{The 10 September 2017 Event} \label{sec3:2017}

During the testbed exercise, we first simulated the solar radiation storm condition based on the 10 September 2017 SEP event. In the following, we describe how we prepared the steady-state solar wind background and simulated the CME and SEP fluxes on-site during the SWPT exercise. 
  
  \subsection{Steady-State Solar Wind Simulation} \label{sec3.2:sw2017}
    
    \begin{figure*}[htp!]
    \centering{
    \includegraphics[width=1.0\textwidth]{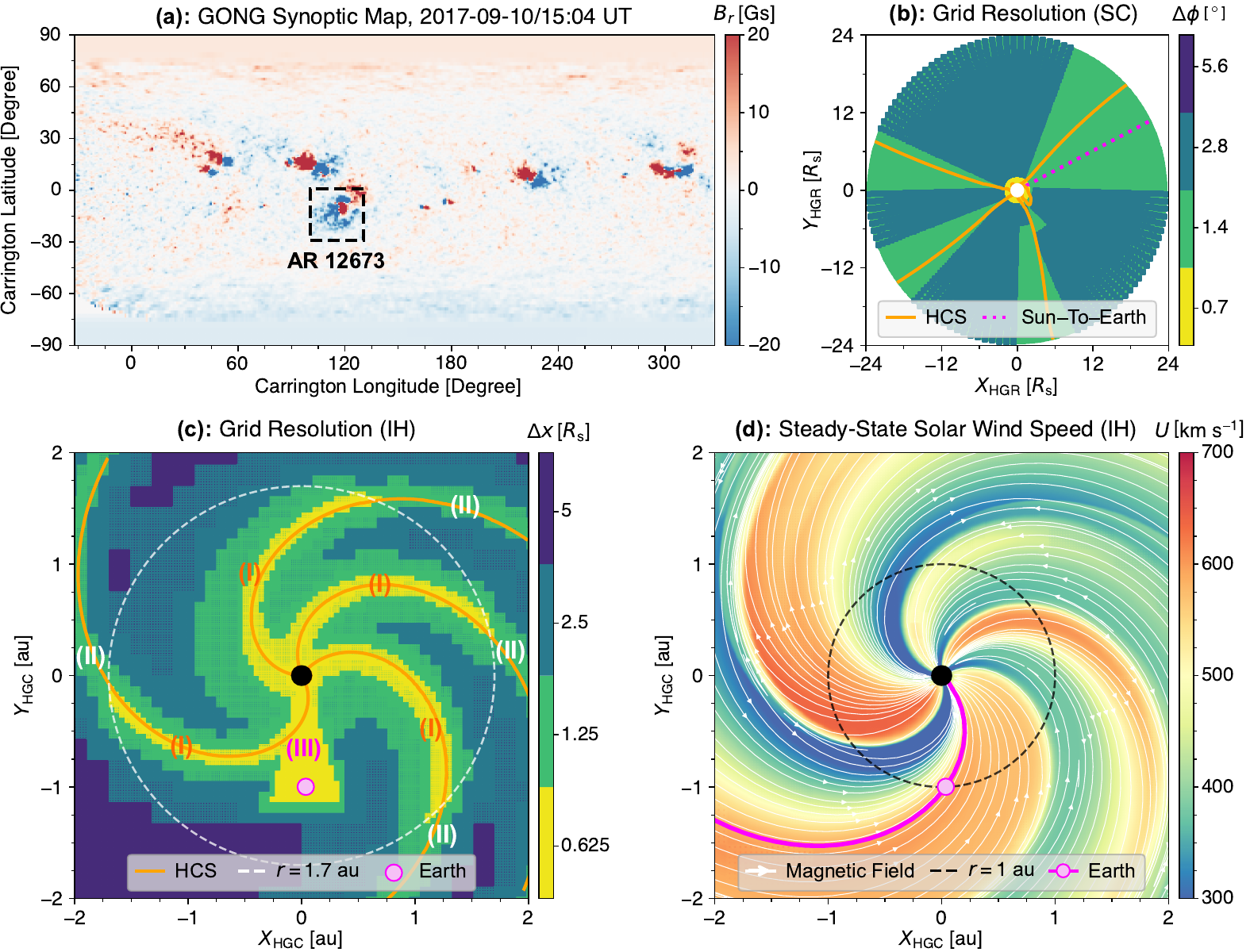}}
    \caption{The input photospheric \textBoldmod{magnetogram}{synoptic map} and the steady-state solar wind solutions for the 10 September 2017 event. 
    (a) Input GONG \textBoldmod{magnetogram}{synoptic map} \textBoldmod{as of}{including observations up to} 15:04 UT on 10 September 2017, with the black dashed box marking the parent AR (NOAA AR 12673) for this event. 
    (b) Angular grid resolution ($\Delta\phi$) of the steady-state simulation in the SC ecliptic plane in heliographic rotating (HGR) coordinates. The white solid circle at the center denotes the inner boundary of the SC domain at a heliocentric distance ($r$) of 1.1 $R_\mathrm{s}$. The HCS is plotted as orange lines, and the Sun-to-Earth connection line is plotted as a magenta dashed line. 
    (c) Mesh size ($\Delta x$) of the steady-state simulation in the IH ecliptic plane in Carrington heliographic (HGC) coordinates. The black solid circle at the center denotes the inner boundary of the IH domain at $r$ = 20 $R_\mathrm{s}$, and the white dashed circle marks the IH shell at $r$ = 1.7 au. The HCS is marked as orange lines. Multiple AMR criteria are indicated as: (I) HCS refinement within the IH shell, (II) HCS refinement beyond the IH shell, and (III) Earth-directed cone refinement. 
    (d) Steady-state solar wind speed in the IH ecliptic plane in HGC coordinates. Multiple white curves with arrows indicate magnetic field lines, with the Earth-connected field line highlighted in magenta. The black solid and dashed circles represent heliocentric distances of 20 $R_\mathrm{s}$ and 1 au, respectively.} \label{fig:steady201709}
    \end{figure*}
    
  The steady-state solar wind simulation was prepared in advance of the SWPT exercise. For the 10 September 2017 event, we used the hourly updated, zero-point corrected synoptic \textBoldmod{magnetogram}{map} from the Global Oscillation Network Group of the National Solar Observatory \cite<NSO/GONG,>[]{harvey1996global, hill2018global}. Figure \ref{fig:steady201709}(a) shows the input GONG \textBoldmod{magnetogram}{synoptic map} \textBoldmod{recorded at}{incorporating magnetograms up to} 15:04 \textBoldmod{UT}{universal time (UT)} on 10 September 2017, right before the solar eruptive event. This \textBoldmod{magnetogram}{synoptic map} has a resolution of 1$^\circ$ in longitude and latitude, corresponding to a $360 \times 180$ grid covering the entire solar surface. We increase the $B_r$ magnitude in its weak-field regions using Equation (1) of \citeA{liu2025physics}. Driven by the processed \textBoldmod{magnetogram}{synoptic map} and a Poynting flux parameter of 0.55 $\mathrm{MW\; m^{-2}\; T^{-1}}$, the stream-aligned AWSoM-R was run in advance to calculate the steady-state solar wind parameters for a period of 27 days centered at this time. \textBoldadd{For this event, the AWSoM-R steady-state solution required $\sim$3 hours of simulation time on 25 Intel Cascade Lake nodes of the NASA Pleiades supercomputer, each with two 20-core Intel Xeon Gold 6248 processors, for a total of 1,000 central processing unit (CPU) cores. This timing highlights the operational motivation for preparing the background solar wind in advance. This preparation is also briefly discussed and summarized below in Section \ref{sec6:sumcon}.}

  In this simulation, we use a block-adaptive 3D spherical grid in the solar corona (SC) domain (1.1–24 solar radii, $R_\mathrm{s}$) and a block-adaptive Cartesian cubic grid in the inner heliosphere (IH) domain (20–650 $R_\mathrm{s}$) that encloses the SC domain, with an overlapping buffer grid between 20 and 24 $R_\mathrm{s}$ that couples the SC and IH solutions. 
  % % The computational domain in SC is composed of grid blocks of $6 \times 8 \times 8$ cells (\textit{control volumes}), with a radially stretched grid constructed by adopting $\ln r$ as the coordinate rather than $r$. 
  % % In the SC domain, the minimal radial mesh size is about $0.02\;R_\mathrm{s}$ near the Sun, to resolve the steep density and temperature gradients in the low corona, and the maximal radial mesh size reaches $\sim$0.8$\;R_\mathrm{s}$ at its upper boundary. 
  In \textBoldadd{the} SC, the angular resolution is initially set as 1.4 degrees. Within $r$ = 1.7 $R_\mathrm{s}$, the angular resolution is increased by one level to 0.7 degrees (see the grid resolution in yellow near the Sun in Figure \ref{fig:steady201709}(b)), while it is coarsened by one level to 2.8 degrees outside of this radial range. 
  % The IH domain encloses the spherical domain of SC%and consists of $8 \times 8 \times 8$ grid blocks
  % , with its mesh size ranging from $\sim0.3\;R_\mathrm{s}$ near the inner boundary to $\sim20\;R_\mathrm{s}$ near the upper boundary. 
  Adaptive mesh refinement \cite<AMR,>[]{gombosi2003adaptive, toth2012adaptive} is applied in both domains to refine the heliospheric current sheet (HCS) and later along the CME propagation direction, ensuring that key coronal and heliospheric structures are adequately resolved. In addition, AMR is applied along an Earth-directed cone to better resolve the solar wind and CME structures most relevant to near-Earth space weather. This grid setup is referred to as Setup 1 for the discussion in Section \ref{sec5.1:timing}. 
  % % The total number of cells is on the order of 2–3 million in SC and about 100 million in IH for these two events in the SWPT exercise. 
  % Note that this is a default SOFIE mesh setup and is used for simulating the 10 September 2017 event. Taking the feedback on-site in the SWPT exercise, we coarsen the grid resolution in SC to simulate the 4 November 2001 event on-site, and finished another simulation using the default mesh setup afterwards for a comparison. More details are explained and discussed below in Sections \ref{sec4:2001} and \ref{sec5:discuss}. 
  
  Figure \ref{fig:steady201709}(b)(c) illustrates the grid resolution in the ecliptic plane within the SC and IH domains, respectively, showing regions of enhanced resolution applied to the HCS, the Earth-directed cone, and specific radial shells of $r$ = 1.7 $R_\mathrm{s}$ in SC and 1.7 au in IH. %This configuration ensures more accurate modeling of the magnetic field topology and solar wind structures relevant for CME and SEP propagation. %In the SC domain, we used the default grid resolution as described in Section \ref{sec2.4:simusetup} for this event. 
  In Figure \ref{fig:steady201709}(d), we show the simulated steady-state solar wind speed in the solar ecliptic plane in IH, plotted with Earth and its connected magnetic field line. We also plot other magnetic field lines as white curves with arrows, demonstrating the alignment between the magnetic field and the solar wind plasma stream \cite<see also>[]{sokolov2022stream, wraback2024simulating, chen2025evidence}.

  \subsection{CME Generation and Propagation} \label{sec3.3:cme2017}

    \begin{figure*}[ht!]
    \centering{
    \includegraphics[width=1.0\textwidth]{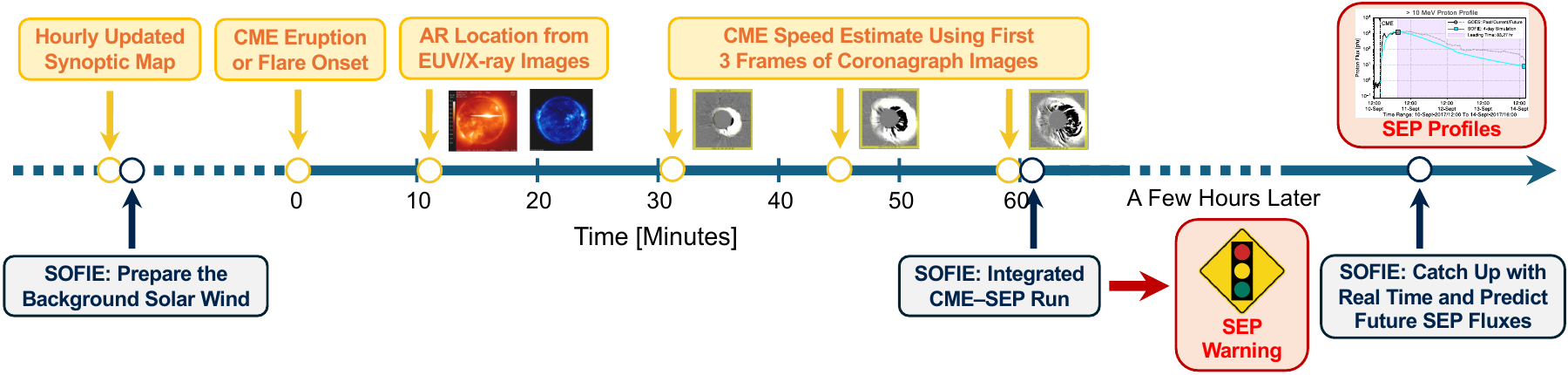}}
    \caption{\textBoldadd{Schematic timeline of the SOFIE workflow under simulated operational conditions for one SEP event. The observational inputs of SOFIE, as shown in Figure \ref{fig:sofie}, and their corresponding available time are illustrated in yellow boxes above the time axis. 
    Key steps for modeling an SEP event by SOFIE in the SWPT exercise, including preparation of the background solar wind and the integrated CME–SEP simulation under simulated operational conditions, are illustrated in blue boxes below the time axis. The resulting SEP flux profile predictions and warnings, illustrated in red boxes, are issued typically a few hours after the eruption, which varies from event to event.}} \label{fig:wflow}
    \end{figure*}

  After preparing steady-state solar wind solutions, we awaited information about the AR location and CME speed during the testbed exercise. 
  At 15:35 UT on 10 September 2017, an X-ray flare was first observed by SWPC forecasters, originating from NOAA AR 12673 at Stonyhurst heliographic latitude $-12^{\circ}$ and longitude $+85^{\circ}$ (S12W85) as viewed from Earth. The flare peaked at $\sim$15:52 UT with an intensity corresponding to the X8.2 class. Shortly afterwards, at 16:00 UT, the associated CME first appeared in the SOHO/Large Angle and Spectrometric Coronagraph \cite<LASCO, described in>[]{brueckner1995lasco} images. During the exercise, we waited until three consecutive LASCO frames were available, allowing a reasonably \textBoldmod{accurate}{confident} estimate of the CME speed, reported as 2650 km s$^{-1}$ by M2M SWAO analysts using the SWPC-CAT. 
  Although \textBoldmod{M2M's}{M2M SWAO analysts'} CME speed determinations were used as model inputs in this study, this selection was made for consistency within the collaborative nature of this exercise but does not represent the formal operational workflow for real-time CME forecasting, \textBoldmod{since SWPC's speed estimates were also available}{in which CME speeds are normally provided by SWPC forecasters}. % nor did the choice of CME speed from SWPC or M2M SWAO greatly affect the SOFIE results
  The current data pipeline in NOAA/SWPC typically incurs a $\sim$10-minute delay to determine the source AR location, depending on the imagery cadence, and it took about 1 hour to obtain the CME speed after the CME eruption. \textBoldadd{Details about this 1-hour delay are discussed in Section \ref{sec6.1:latency}.} 
  \textBoldadd{Figure \ref{fig:wflow} provides a schematic timeline summarizing the operational workflow for each event in the SWPT exercise, particularly during the early onset phase, including the CME eruption time, data availability, model initiation, and the forecast window. Details about the runtime and efficiency of SOFIE throughout the SEP event are described later in Sections \ref{sec3.4:sep2017}, \ref{sec4.4:sep2001}, and \ref{sec5.1:timing}.} 
  
\begin{figure*}[htp!]
    \centering{
    \includegraphics[width=0.95\textwidth]{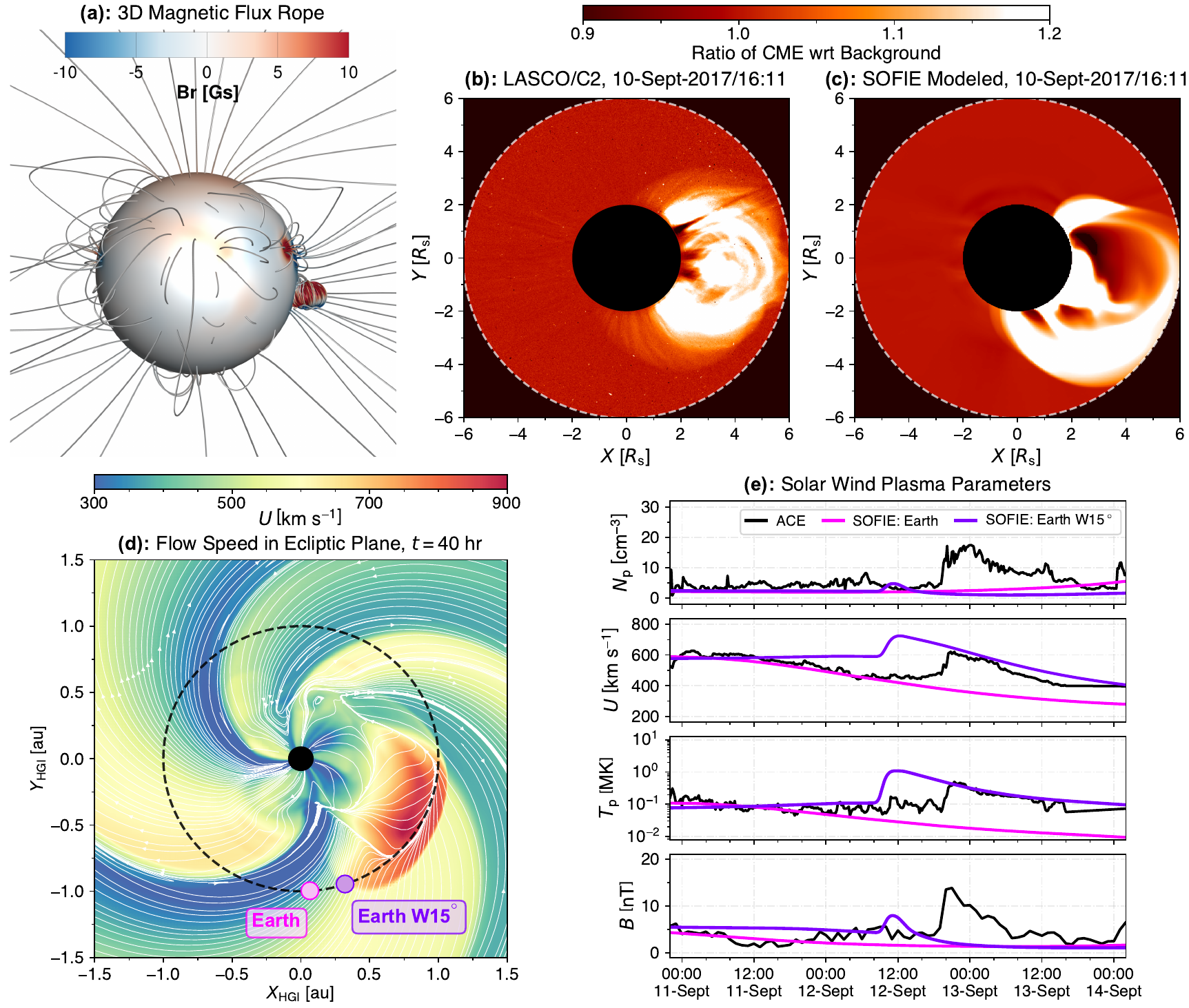}}
    \caption{CME simulation results during the 10 September 2017 event. 
    Panel (a) shows the initial 3D CME flux rope inserted above the parent AR at the inner boundary of \textBoldadd{the} SC \textBoldadd{domain} ($r$ = 1.1 $R_\mathrm{s}$), viewed from Earth and colored by the radial magnetic field strength ($B_r$). 
    Panels (b)–(c) show the LASCO/C2 observation and SOFIE-synthesized WL images at 16:11 UT on 10 September 2017. The color scale indicates the relative change in WL total brightness, expressed as the ratio of the CME signal to the background solar wind before the eruption. In both panels, the field of view is limited within $r$ = 6 $R_\mathrm{s}$, with the central black solid circle marking an occultation disk at $r$ = 2 $R_\mathrm{s}$. 
    Panel (d) shows the flow speed in the IH ecliptic plane in heliographic inertial (HGI) coordinates, 40 hours after the eruption, when the ICME flank approaches Earth. Multiple magnetic field lines are marked as white curves with arrows, and a black dashed circle denotes the heliocentric distance of 1 au. The magenta and purple scatter points denote Earth and its 15$^\circ$ westward position in the ecliptic plane, respectively. 
    Panel (e) shows the modeled solar wind plasma parameters at Earth (\textBoldadd{in} magenta) and its 15$^\circ$ westward position in the ecliptic plane (\textBoldadd{in} purple), together with ACE observations at Earth (\textBoldadd{in} black). Parameters shown from top to bottom are the plasma number density ($N_\mathrm{p}$), solar wind speed ($U$), plasma temperature ($T_\mathrm{p}$), and magnetic field strength ($B$). 
    } \label{fig:cmeshk201709}
    \end{figure*}

  Then, we used the \textBoldmod{magnetogram}{synoptic map}, AR location, and CME speed as inputs (see Table \ref{tab:param}) for EEGGL to parameterize the CME flux rope. As shown in Figure \ref{fig:cmeshk201709}(a), we inserted the resulting force-imbalanced spheromak-type magnetic flux rope and its entrained plasma on top of the parent AR (NOAA AR 12673 for this event). 
  For both events in the SWPT exercise, SOFIE was run \textBoldmod{on 25 Intel Cascade Lake nodes of the NASA Pleiades supercomputer, each with two 20-core Intel Xeon Gold 6248 processors, for a total of}{on-site using} 1,000 CPU cores \textBoldadd{of Intel Cascade Lake nodes}. %This resource configuration overall provides optimal parallelization efficiency for our current grid setup. 
      
  The modeled synthetic white-light (WL) images are compared with LASCO observations at 16:11 UT on 10 September 2017, as shown in Figure \ref{fig:cmeshk201709}(b)(c). These images represent the region from 2 to 6 $R_\mathrm{s}$ within the LASCO/C2 field of view, with the inner black circle denoting an occultation disk at 2 $R_\mathrm{s}$. The color scale represents the relative change in WL total brightness with respect to the steady-state solar wind background. 
  Here, both the observation and the simulation show the CME propagating westward, with the bright leading front reaching a comparable distance of $\sim$6 $R_\mathrm{s}$ in the plane of the sky at this time. The simulated results also clearly exhibit the classic three-part CME morphology \cite<e.g.,>[]{illing1985observation}. 

  As the CME propagates outward and evolves into an interplanetary CME (ICME), the shock flank approaches Earth about 40 hours after the eruption. Figure \ref{fig:cmeshk201709}(d) shows the solar wind speed with magnetic field lines in the IH ecliptic plane at this time. In the simulation, the ICME flank extends toward Earth but does not directly impact it, likely because of the presence of the HCS acting as an obstacle and Earth's location $\sim$90$^\circ$ in longitude away from the source AR. To better examine the \textBoldmod{flank properties}{solar wind properties at the flank}, we consider a reference position 15$^\circ$ westward of Earth in the ecliptic plane. Figure \ref{fig:cmeshk201709}(e) compares the simulated solar wind plasma parameters at these two locations with the ACE \textit{in situ} observations at Earth. 
  In observations, the shock arrived at $\sim$20:00 UT on 12 September, whereas the simulated ICME does not reach Earth. At the 15$^\circ$ westward position, the simulated ICME shock front arrived 8–10 hours earlier than observed by \textBoldmod{ACR}{ACE}, with a comparable solar wind speed and ion temperature, but lower plasma density and magnetic field strength. % These differences may result from the flank geometry and complicated interactions between the CME and the background HCS and SIR, leading to some noticeable deviations in the modeled solar wind parameters for this ICME at 1 AU.  
  In our simulation, we find that the CME/ICME is significantly deflected as a result of interaction with the HCS and the stream interaction region in the background solar wind. 
  We also note that the choice of both steady-state model and flux-rope parameters can impact the simulated plasma and magnetic properties of the CME \cite<e.g.,>[]{jivani2023global, huang2024solar, chen2025decent}. However, in this study for simulated real-time operations, we do not explore varying the model parameters.

  \subsection{SEP Distribution and Fluxes} \label{sec3.4:sep2017}
  
    \begin{figure*}[htp!]
    \centering{
    \includegraphics[width=0.95\textwidth]{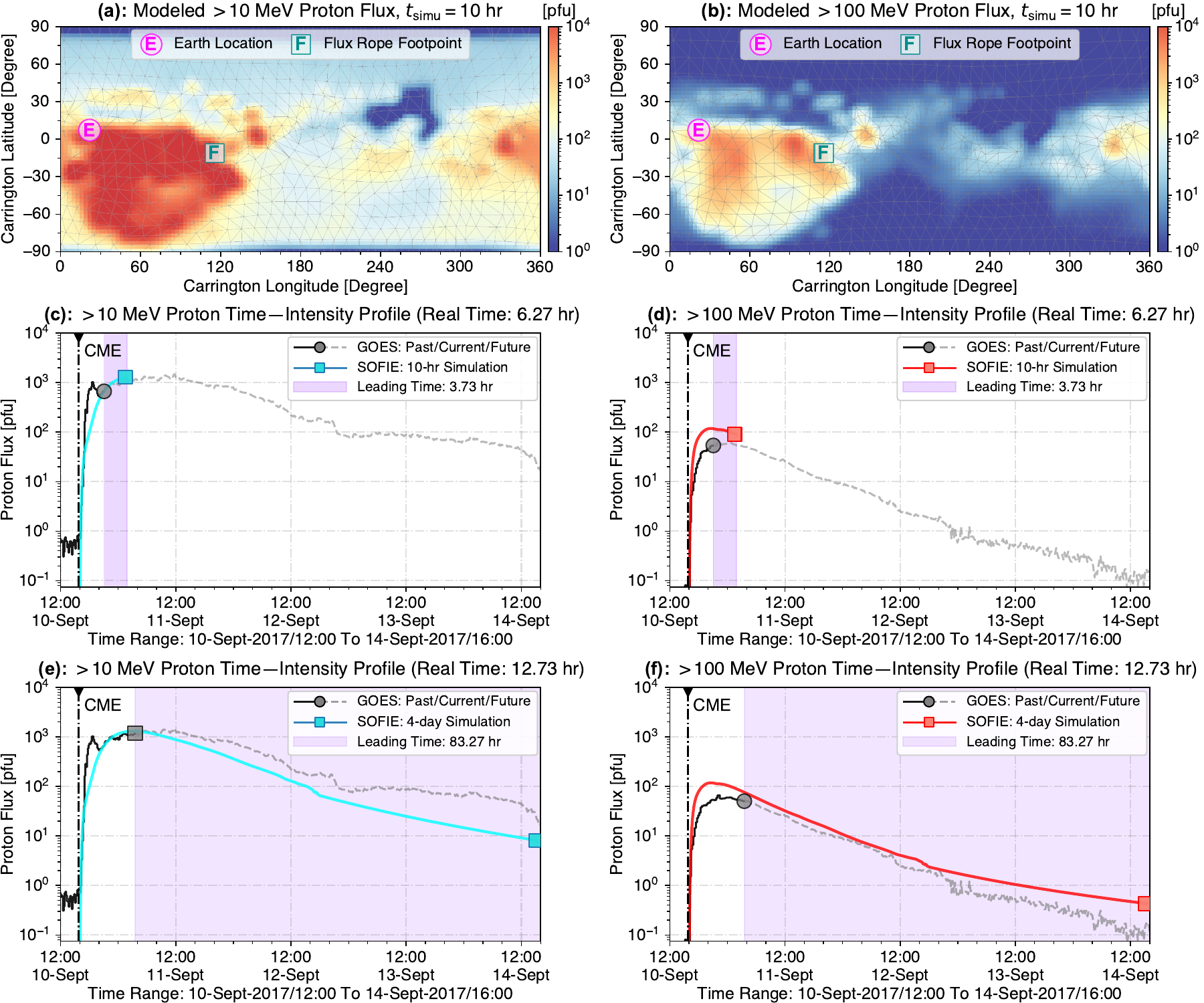}}
    \caption{Distribution of energetic protons in the 10 September 2017 event. 
    Panels (a) and (b) show SOFIE-modeled fluxes of $>$10 MeV and $>$100 MeV protons, respectively, on a logarithmic scale on the 1 au sphere, 10 hours after the eruption. In each panel, the $x$- and $y$-axes are Carrington longitude and latitude, respectively. The thin gray vertices correspond to the field line intersections with the 1 au sphere, and the edges indicate the triangulation skeleton constructed via the Delaunay triangulation approach \cite{delaunay1934sphere, lee1980two}. The Earth's location is marked as ``E" in a magenta circle, and the CME flux rope footpoint on the solar surface is marked as ``F" in a green square. 
    Panels (c) and (d) show the $>$10 MeV and $>$100 MeV proton time--intensity profiles at Earth, respectively, 6.27 real-time hours after the eruption, corresponding to the time when SOFIE predicts 10-hour proton fluxes. In each panel, GOES observations are plotted as a black curve, followed by a dark gray circle and a light gray curve indicating past, current, and future fluxes, respectively. SOFIE results are plotted as cyan and red curves with square markers, representing the $>$10 MeV and $>$100 MeV energy channels, respectively. A dashed-dotted vertical line represents the CME eruption time, and a vertical purple band marks the prediction leading time of 3.73 hours, also indicated in the legend. 
    Panels (e) and (f) are similar to panels (c) and (d), but 12.73 real-time hours after the eruption, when SOFIE predicts 4-day proton fluxes, yielding a leading time of 83.27 hours. 
    } \label{fig:sep201709}
    \end{figure*}
  
  As the AWSoM-R MHD simulation was initiated on-site during the SWPT exercise, M-FLAMPA simultaneously solved the distribution function of energetic protons along magnetic field lines, providing both spatial and temporal characteristics about the SEP distribution. 
  Figure \ref{fig:sep201709}(a)(b) illustrates the modeled $>$10 MeV and $>$100 MeV proton fluxes on the 1 au sphere 10 hours after the launch of the CME flux rope. In each panel, the Earth's location and the CME flux rope footpoint on the Sun's surface are marked as ``E" in a magenta circle and ``F" in a green square, respectively. 
  We can find obvious longitudinal and latitudinal variations of the energetic proton fluxes. In the $>$10 MeV channel (Figure \ref{fig:sep201709}(a)), proton fluxes can reach as high as $10^{4}$ pfu, and in the $>$100 MeV channel (Figure \ref{fig:sep201709}(b)) they can reach $\sim$10$^{3}$ pfu, concentrated on the 1 au sphere primarily within $0^{\circ}$–$120^{\circ}$ east of the flux rope footpoint. In contrast, on the 1 au sphere, regions $45^{\circ}$–$180^{\circ}$ west of the flux rope, corresponding to the backside of CME propagation, barely exhibit SEPs, with fluxes 2–4 orders of magnitude lower. 
  \textBoldadd{These spatial variations illustrate the critical role of large-scale magnetic connectivity between the CME-driven shock and the observer: regions with favorable connectivity experience substantially higher SEP fluxes, while poorly connected regions receive minimal fluxes \cite<see also, e.g.,>{cane1988role, lario2013longitudinal, lario2017link}.} 
  
  Figure \ref{fig:sep201709}(c)–(d) shows the modeled and observed proton time--intensity profiles at Earth, 6.27 real-time hours after the eruption, when SOFIE provides 10-hour proton flux predictions. Here, the real time is counted since the flare onset and incorporates the $\sim$1-hour delay for obtaining CME information as SOFIE inputs. The GOES observations are plotted as black and gray curves, representing the past, current, and future fluxes, respectively. The SOFIE-predicted $>$10 MeV and $>$100 MeV proton fluxes are shown as cyan and red curves with square markers. 
  At this moment, the GOES observed proton fluxes were still increasing toward their peak, while SOFIE already reproduced comparable onset and peak fluxes within a factor of 2, with a prediction leading time of 3.73 hours\textBoldadd{, defined hereafter as the difference between the simulation time and the real time, and also used as a model performance metric}. 
  
  The \textBoldmod{time profiles for}{SEP time–intensity profiles at Earth from} the completed 4-day SOFIE simulation are presented in Figure \ref{fig:sep201709}(e)–(f) using the same plot style. Overall, the modeled profiles show very good agreement with the GOES observations, from the onset and peak to decay, and in terms of time and fluxes. 
  Specifically, in this event, a magnetic cloud has been reported to pass Earth $\sim$2 days after the eruption \cite<see Figure \ref{fig:cmeshk201709}(e) and>[]{guo2018modeling, luhmann2018shock}. For SEPs observed by GOES, this interval corresponds to a small plateau in the $>$10 MeV channel and a modest drop thereafter in both the $>$10 MeV and $>$100 MeV channels around 20:00 UT on 12 September 2017. In the simulation, a similar decrease was reproduced but occurred about 4–5 hours earlier than observed, mainly because the modeled ICME flank approached Earth earlier (see Figure \ref{fig:cmeshk201709}(d)(e))\textBoldadd{, although the CME/ICME arrival at Earth is not perfectly captured in the \textit{in situ} solar wind plasma comparison}. Although the \textBoldadd{simulated} ICME did not reach Earth directly, the connectivity between Earth and the CME/ICME-driven shock could have changed. Nonetheless, this timing discrepancy is relatively minor compared to the $\gtrsim$4-day duration of this event, particularly given that the interplanetary shock did not produce a strong \textBoldmod{ESP enhancement}{enhancement of energetic storm particles (ESPs)} in this case. 
  % For SEPs observed by GOES, this interval corresponds to a small plateau in the $>$10 MeV energy channel and a modest drop afterwards in both the $>$10 MeV and $>$100 MeV energy channels, at $\sim$20:00 UT on 12 September 2017. The SOFIE simulation reproduces the feature of a similar drop, despite being a few hours earlier than observed, which appears to be comparable given the multi-day timescale of the event. 
  Notably, the 4-day SOFIE simulation completed within 13 hours of real time, at a point when the observed fluxes were still near their peak or just beginning to decay, providing the full forecast profile up to 4 days in advance. 
  
  % The computational efficiency of SOFIE in Figure \ref{fig:timing} also reflects the changing physical regime throughout the event. 
  % Early in each simulation, when the flux rope resides in the SC domain, the cell size is small (on the order of $10^{-2}$–$10^{-1}\; R_\mathrm{s}$) and the fast-mode magnetosonic speed is high (hundreds of km s$^{-1}$), requiring small global time steps (on the order of $10^{-1}$ s) in the MHD model. We set a coupling time of 2 minutes between SC and IH, allowing frequent parameter exchanges but also increasing the computational demand. In SP/M-FLAMPA, suprathermal particles are accelerated through the DSA mechansim near the shock front, also requiring small time steps (typically on the order of $10^{-1}$–$10^{0}$ s) that add additional cost. 
  % After 12 hours of the onset, the CME flux rope has fully left the SC domain, so the SC-IH coupling frequency is reduced to 4 hours. The global time step for solving the MHD equations then grows to minutes. In addition, the time step in SP/M-FLAMPA increases to 2 minutes for most field lines, as particle transport dominates over acceleration. Together, all these factors substantially reduce the computational expense at later times. 
  
  During the SWPT exercise, we were allocated 4–6 hours \textBoldadd{to test the operational workflow} for each event. Within this limited time, SOFIE was able to provide only the first several hours of predictions. \textBoldadd{Specifically, SOFIE predicted the threshold crossing time for $>$10 MeV, $>$10 pfu 1.46 hours after the flare onset (1 hour for the input delay, plus 0.46 hours of running time), and for $>$100 MeV, $>$1 pfu 1.26 hours after the flare onset. However, the GOES observation reported corresponding threshold crossing times 57 and 27 minutes after the flare onset, respectively, indicating that the earliest SEP threshold warnings from SOFIE lagged the first observed threshold crossing by $\sim$1 hour in real time. For this event simulation, we also find that SOFIE caught up with the real time after 3.21 hours, again suggesting a delay of the earliest forecast of SEP flux warning.} 
  \textBoldmod{We}{Here, we} note that the most time-consuming part of the simulation lies in the SC domain, where the grid size is small (on the order of $10^{-2}$–$10^{-1}\; R_\mathrm{s}$) and the MHD characteristic wave speeds are high (hundreds of km s$^{-1}$), thereby requiring very small global time steps (on the order of $10^{-1}$ s). As the CME flux rope propagates outward from \textBoldadd{the} SC \textBoldadd{domain} into \textBoldadd{the} IH, the simulation progressively becomes faster due to coarser grid spacing and lower characteristic speeds \textBoldadd{(see also Figure \ref{fig:timing} and associated texts for the change in runtime speed)}. 
  In-depth discussions with forecasters, operators, and analysts on-site in the SWPT exercise led to a practical suggestion and test, which is to coarsen the grid in \textBoldadd{the} SC \textBoldadd{domain} by a factor of two and evaluate the impact on both accuracy and speed in the next event simulation.

\section{The 4 November 2001 Event} \label{sec4:2001}

On the second day of the SWPT exercise, we simulated the solar eruptive event on 4 November 2001. We describe the approach for preparation, simulation and SOFIE results for this event similar to Section \ref{sec3:2017}.

  \subsection{Steady-State Solar Wind Simulation} \label{sec4.2:sw2001}
  
\begin{figure*}[ht!]
    \centering{
    \includegraphics[width=1.0\textwidth]{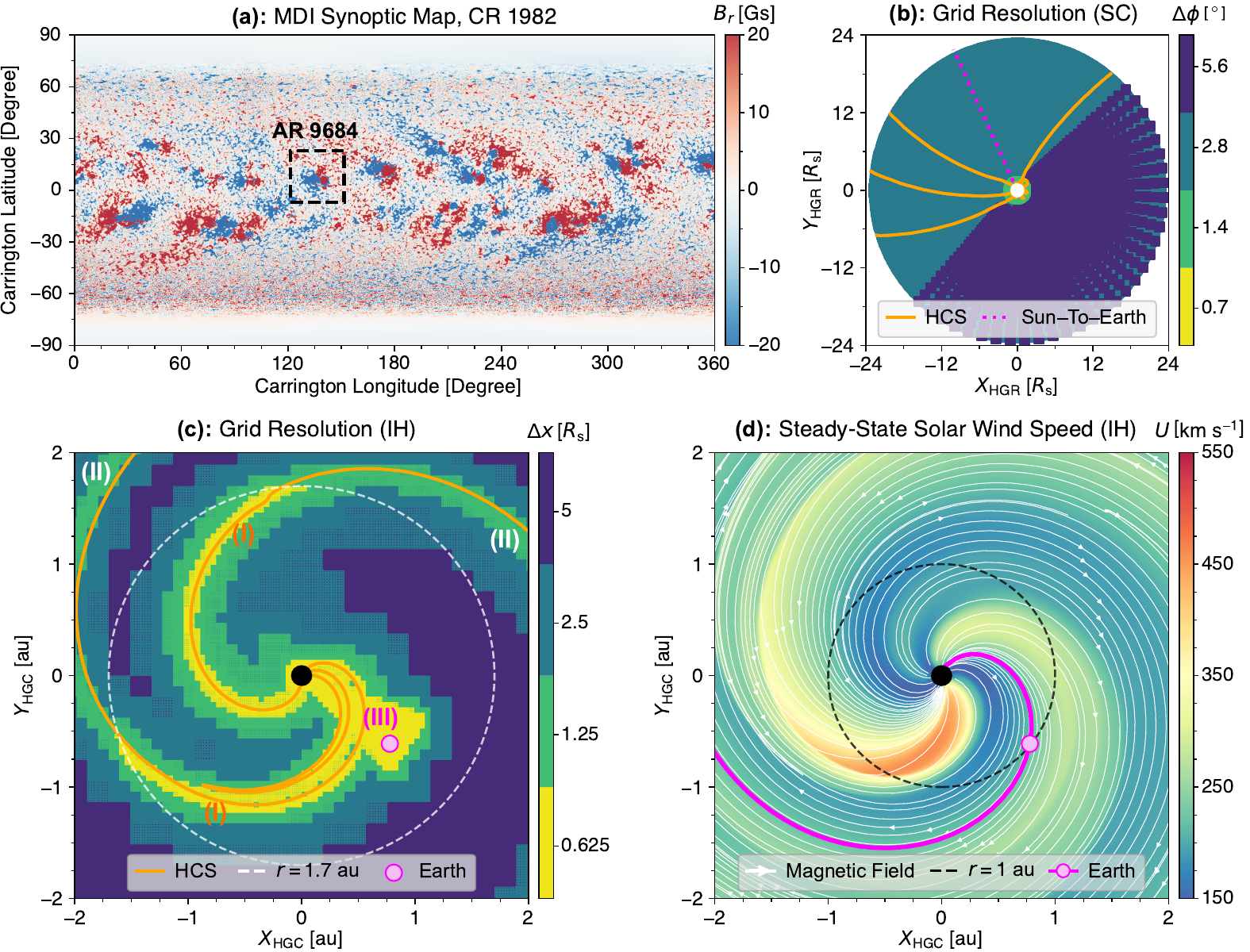}}
    \caption{The input photospheric \textBoldmod{magnetogram}{synoptic map}, the angular resolution of the grid in \textBoldadd{the} SC, the mesh size in \textBoldadd{the} IH, and the steady-state solar wind solutions in \textBoldadd{the} IH for the 4 November 2001 event, shown in panels (a)–(d), respectively, with the same plot style as Figure \ref{fig:steady201709} except for adjusted plotting ranges and ticks.} \label{fig:steady200111}
    \end{figure*}

  With the feedback from the first event simulation (see the end of Section \ref{sec3.4:sep2017}), we coarsened the grid in SC by a factor of two (referred to as Setup 2 hereafter) when we prepared the background solar wind. Since GONG magnetograms are only accessible after September 2006, we used a zero-point corrected, Carrington rotation (CR) synoptic map based on the SOHO/Michelson Doppler Imager \cite<MDI,>[]{scherrer1995solar} observations. 
  Figure \ref{fig:steady200111}(a) shows the input MDI \textBoldmod{magnetogram}{synoptic map}, integrated over CR 1982 from 17 October to 13 November 2001. It has a resolution of 0.1$^\circ$ in longitude and 0.125$^\circ$ in latitude, corresponding to a $3600 \times 1440$ grid covering the entire solar surface. We remap it to a $360 \times 180$ grid with 1$^\circ$ resolution in both longitude and latitude and amplify $B_r$ in the weak-field regions using Equation (1) of \citeA{liu2025physics}. 
  The global magnetic field strength during this period is notably stronger than that in September 2017 (see Figure \ref{fig:steady201709}(a)), consistent with the fact that this SEP event occurs near the solar cycle 23 maximum. 
  We prepared the steady-state solar wind plasma over this CR using a Poynting flux parameter of 0.2 $\mathrm{MW\; m^{-2}\; T^{-1}}$ in AWSoM-R\textBoldadd{, which required approximately 1–1.5 hours of simulation time on 1,000 CPU cores even for this coarser grid setup}.
  
  In Figure \ref{fig:steady200111}(b)(c), we show the grid resolution in the SC and IH domains, respectively. Compared with the mesh setup in the previous event (Figure \ref{fig:steady201709}(b)(c)), the overall angular resolution in \textBoldadd{the} SC is coarsened by a factor of 2, and the mesh size for the heliospheric structures in \textBoldadd{the} IH remains the same. 
  In Figure \ref{fig:steady200111}(d), we plot the simulated steady-state solar wind speed in the ecliptic plane in the IH domain using the same style as in Figure \ref{fig:steady201709}(d). The results again demonstrate the resolved heliospheric structure and the expected alignment between the magnetic field and the solar wind flow, and show the global ambient solar wind background into which the CME and SEPs are subsequently simulated to propagate.

  \subsection{CME Generation and Propagation} \label{sec4.3:cme2001}
  
\begin{figure*}[ht!]
    \centering{
    \includegraphics[width=0.95\textwidth]{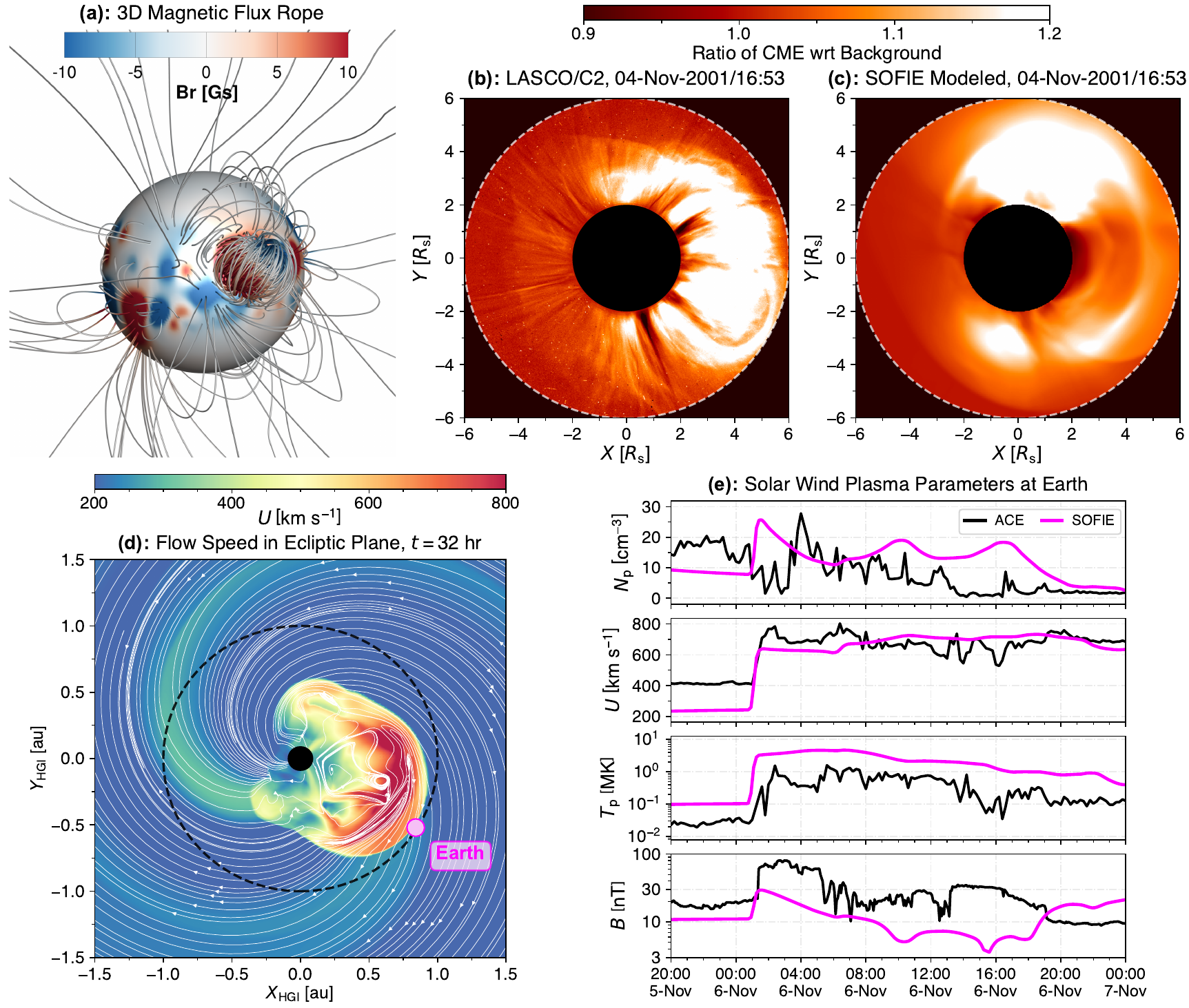}}
    \caption{CME generation and propagation results during the 4 November 2001 event, including the initial 3D CME flux rope in panel (a); the observed and simulated LASCO/C2 WL images at 16:53 UT on 4 November 2001 shown in panels (b)–(c); the flow speed in the IH ecliptic plane ecliptic shown in panel (d); and the solar wind plasma parameters at Earth in panel (e), presented in the same plot style as Figure \ref{fig:cmeshk201709}, except for detailed plotting ranges and ticks. 
    % Panel (d) shows the flow speed in the ecliptic plane in heliographic inertial (HGI) coordinates, 32 hours after the eruption, when the ICME is about to pass through Earth. Multiple magnetic field lines are marked as white curves with arrows, and a black dashed circle denotes the heliocentric distance of 1 au. 
    % Panel (e) shows the solar wind plasma parameters at Earth, which are the plasma number density ($N_\mathrm{p}$), solar wind speed ($U$), plasma temperature ($T_\mathrm{p}$) and magnetic field strength ($B$, plotted on a logarithmic scale) shown from the top to the bottom. ACE observations are shown in black, and SOFIE results are shown in magenta. 
    } \label{fig:cmeshk200111} % except for detailed plotting ranges and tickss
    \end{figure*}

  During the testbed exercise, we monitored solar activity using real-time data. At 16:09 UT on 4 November 2001, SWPC forecasters reported an X-ray flare erupting from NOAA AR 9684 at N05W24 as viewed from Earth, which peaked around 16:18 UT with an X1.0-class intensity. Shortly after, a fast-halo CME was first detected by LASCO/C2 at 16:35 UT. After three consecutive LASCO frames became available, which is $\sim$1 hour after the flare onset, a plane-of-sky CME speed of 1925 km s$^{-1}$ was provided by the M2M SWAO analysts using the SWPC-CAT. 
  Taking the MDI \textBoldmod{magnetogram}{synoptic map}, together with the identified AR location and CME speed listed in Table \ref{tab:param}, we generated the CME flux rope via EEGGL and placed it above the parent AR (see Figure \ref{fig:cmeshk200111}(a)). 
  
  We show the WL images from the LASCO/C2 coronagraph and SOFIE simulations at 16:53 UT on 4 November 2001. The LASCO/C2 observation in Figure \ref{fig:cmeshk200111}(b) shows a halo CME propagating predominantly westward, with its bright core extending northward. The corresponding SOFIE-synthesized WL image in Figure \ref{fig:cmeshk200111}(c) reproduces the overall flux rope structure, also appearing as a westward-directed halo CME that reaches approximately the same radial distance (6 $R_\mathrm{s}$) as observed. 
  % The major discrepancy between these two panels is the enhanced brightness in the northern part of the simulated flux rope compared to the observation. This difference is likely owing to projection effects \cite<e.g.,>[]{temmer2009cme}, since the parent AR lies near the disk center, where even small variations in the LOS integration can lead to noticeable brightness asymmetries. 
  
  As the CME propagates from the SC to the IH domain evolving into an ICME, the CME-driven shock arrives at Earth 32 hours after the eruption. Figure \ref{fig:cmeshk200111}(d) shows the solar wind speed, together with multiple magnetic field lines, in the ecliptic plane at this time. As shown, the ICME is directed toward Earth, with its shock front impacting Earth and a sheath region followed by the magnetic cloud. 
  In Figure \ref{fig:cmeshk200111}(e), we compare the solar wind plasma parameters at Earth from ACE \textit{in situ} observations and SOFIE simulations, which are overall in good agreement. The shock arrival time, $\sim$01:30 UT on 6 November 2001, aligns well between ACE observations and SOFIE simulations. The amplitudes of the solar wind speed, plasma number density, and temperature after the shock arrival also agree well. While ACE observes the peak of the magnetic field amplitude of $\sim$80 nT, SOFIE predicts about 30 nT. This is a typical amplitude for ICME-driven shocks arriving at Earth in AWSoM/EEGGL simulations \cite<e.g.,>[]{manchester2004modeling, jin2017chromosphere, chen2025decent}. We also note several CMEs occurred in the days preceding this event \cite{kuznetsov2003solar, shen2008enhancement}. As this large CME propagates outwards, it can interact or merge with the earlier CMEs/ICMEs, resulting in an enhanced magnetic field amplitude at 1 au as observed by ACE. However, since the present analysis focuses on the major event during the testbed exercise, those preceding CMEs were not included in the simulation, which could have distorted the solar wind background and accounted for the differences in the comparison above.

  \subsection{SEP Distribution and Fluxes} \label{sec4.4:sep2001}
    
\begin{figure*}[ht!]
    \centering{
    \includegraphics[width=0.95\textwidth]{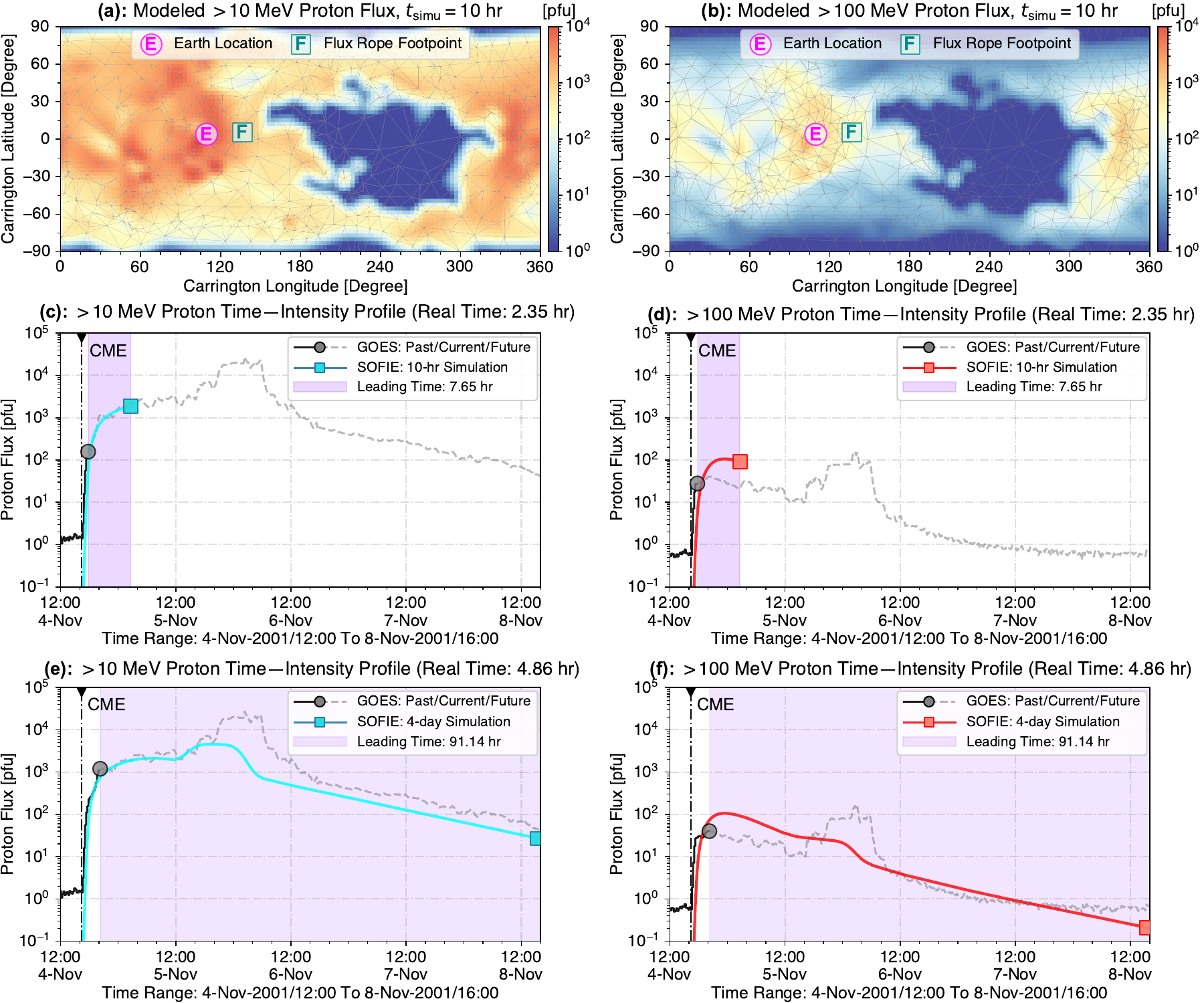}}
    \caption{Distribution of energetic protons during the 4 November 2001 event, including the modeled $>$10 MeV and $>$100 MeV proton fluxes on the 1 au sphere at 10 hours after the eruption shown in panels (a)–(b), and the proton time--intensity profiles shown in panels (c)–(f), plotted in the same style as Figure \ref{fig:sep201709} except for adjusted plotting ranges and ticks.} \label{fig:sep200111}
    \end{figure*}

    As the simulated CME propagates, M-FLAMPA outputs the distribution function of the energetic protons along multiple field lines. Figure \ref{fig:sep200111} illustrates the resulting spatial distribution and temporal evolution of energetic protons during the 4 November 2001 SEP event, following the same plotting pattern as Figure \ref{fig:sep201709}. Similarly, in Figure \ref{fig:sep200111}(a)(b), we plot the modeled proton fluxes at energies $>$10 MeV and $>$100 MeV on the 1 au sphere, 10 hours after the eruption, showing broad variations of the SEP fluxes in longitudes and latitudes. 
    \textBoldadd{These spatial variations again highlight that SEP fluxes are primarily determined by the global magnetic connectivity and shock properties. While local plasma conditions may not be perfectly reproduced in Figure \ref{fig:cmeshk200111}(e), which results in some timing and amplitude offsets during the ESP phase at Earth, the SOFIE model can still provide a physically plausible representation of the global SEP distribution throughout the inner heliosphere (Figure \ref{fig:sep200111}(a)–(b)) and overall reasonable agreement with SEP observations at Earth (Figure \ref{fig:sep200111}(d)–(f)).} 
    
    The $>$10 MeV and $>$100 MeV proton time--intensity profiles at Earth are presented in Figure \ref{fig:sep200111}(c)–(f), following the same plot style as Figure \ref{fig:sep201709}(c)–(f). We show the profiles for 10 and 96 simulated hours after the eruption in panels (c)(d) and (e)(f), corresponding to 2.35 and 4.86 real-time hours after the eruption, respectively. 
    Both observed and simulated proton fluxes rise rapidly after the CME eruption. For the $>$10 MeV channel, SOFIE/M-FLAMPA predicts an onset peak of $\gtrsim$1,000 pfu, in close agreement with the GOES observation of $\sim$3,000 pfu within a factor of 2–3. In the $>$100 MeV channel, SOFIE/M-FLAMPA gives a higher peak of $\sim$100 pfu, compared to the $\sim$40 pfu measured by GOES. 
    \textBoldadd{In this simulation, SOFIE provided the SEP threshold crossing time prediction with a delay of $\lesssim$0.6 hours with respect to the GOES observations for both energy channels (see Setup 2 in Table~\ref{tab:compres}), as the simulation was initiated after CME inputs became available $\sim$1 hour after the eruption. A more detailed assessment of SOFIE’s operational timing and latency is presented in Section~\ref{sec5.1:timing}.} 
    
    % Note that a strong ICME-driven shock propagates toward Earth, accelerating particles locally but over a shorter transport distance to Earth. This results in an increasing proton flux in 1 au observation, but primarily affects the profiles for protons up to tens of MeV \citep[e.g.,][]{cohen2006observations, lario2018flat, lario2019evolution}. 
    % Consequently, the observed $>$10 MeV proton flux shows a gradual increase, while the $>$100 MeV flux declines slightly after the onset phase. In the simulation, the main features of these two energy channels are basically reproduced, although the $>$100 MeV profile maintains a higher plateau after the onset, suggesting that the shock may have remained stronger than expected in interplanetary space. 
    In this event, a strong ICME-driven shock propagates toward Earth, accelerating the particles locally but over a shorter transport distance to Earth. After 12:00 UT on 5 November 2001, we can find a second enhancement in both $>10$ MeV and $>$100 MeV proton fluxes observed by GOES, which is the ESP peak and coincides with the ICME shock passage later at $\sim$01:30 UT on 6 November. In the simulation, the main features of these two energy channels are basically reproduced, although both profiles maintain lower plateaus. This difference implies that the shock may have remained stronger than predicted in interplanetary space, and that the diffusive nature of the simulation (especially under the reduced spatial resolution) and the preceding CMEs/ICMEs not included in our SWPT simulations could have played a role in shaping the ESP phase. 
    
    Overall, the SOFIE/M-FLAMPA simulation captures the main characteristics of this intense SEP event, especially the amplitude and timing of the SEP onset and ESP peak, despite some detailed deviations in the time--intensity profile. In particular, using a coarser grid setup in the SC domain for this event simulation, we can predict the 4-day profile in 5 hours of real time, which showcases the operational usefulness of the SOFIE model. 
    % These results demonstrate SOFIE's capability to predict large SEP events, capturing both the onset and ESP phases, and highlight its potential for application to other historical and future SEP events. 

\section{Discussion} \label{sec5:discuss}

  \subsection{\textBoldadd{Coronagraph Latency and Forecast Timeliness Considerations}} \label{sec6.1:latency}

      \textBoldadd{For the two SEP events simulated during the SWPT exercise, we adopted an input latency of $\sim$1 hour for the CME speed estimation as the SOFIE input (see Figure \ref{fig:wflow}). In fact, this $\sim$1-hour delay accounts for the latency of $\sim$20–30 minutes for the availability of the first coronagraph images used to estimate CME kinematics under nominal conditions, together with the additional time required to obtain three consecutive images for the CME speed estimate. 
      % This latency consideration is consistent with the current availability of coronagraph data from the GOES-19 satellite launched in 2024% \cite{giri2025review} % This may be read to suggest CCOR-1 latency is comparable to LASCO and thus minimizes the significance of the CCOR instruments. In truth, the introduction of CCOR-1 as the first operational coronagraph is a major advancement in SWx forecasting. It affords a reliable image cadence streamed to the forecast office every (<)30min. Whereas, prior coronagraphs may have offered a nominal latency of ~1hour, the reality has been that delays of many hours were quite common. 
      % , whose Compact Coronagraph 1 (CCOR-1) instrument
      While this latency assumption might represent an optimistic real-world case during the SWPT exercise, it is consistent with current operational practices. In particular, the GOES-19 Compact Coronagraph 1 (CCOR-1) instrument \cite<\url{https://ccor.nrl.navy.mil/},>[]{thernisien2025ccor}, launched in 2024, represents a major advancement in space weather predictions as the first operational coronagraph, which affords a latency of $\sim$30 minutes from image capture until availability to SWPC forecasters and a cadence of 15 minutes. 
      While this 1-hour input delay corresponds to the historic nominal operational scenario (see \url{https://lasco-www.nrl.navy.mil/index.php?p=content\%2Frealtime} and \url{https://www.swpc.noaa.gov/}), we acknowledge that, historically, latencies in real-world varied depending on data availability and quality control procedures, and that additional processing or coordination steps introduced additional delays \cite<e.g.,>[]{posner2014main}. With the deployment of CCOR-1, these latencies have been drastically reduced reliably to $\lesssim$30 minutes.} 

  \subsection{Impacts of Grid Resolution on SEP Prediction Accuracy and Timing} \label{sec5.1:timing}
  
  While SOFIE was run on a coarser grid setup for the SC domain (Setup 2) described in Section \ref{sec3.4:sep2017}, we also used the default grid configuration (Setup 1), as described in Section \ref{sec3.2:sw2017}, to run SOFIE after the SWPT exercise. In addition, on-site in the SWPT exercise, interactive discussions were held among the CLEAR team and SWPC forecasters, SRAG console operators, and M2M SWAO analysts, suggesting a grid setup that maintains the coarser grid resolution for the background as in Setup 2 and keeps the default resolution for the HCS, as well as CME-faced and Earth-oriented cones (Setup 3). We then used these grid setups to run SOFIE and conducted a comparative study in terms of the SEP prediction accuracy and timing. 
  
    \begin{figure*}[ht!]
    \centering{
    \includegraphics[width=0.95\textwidth]{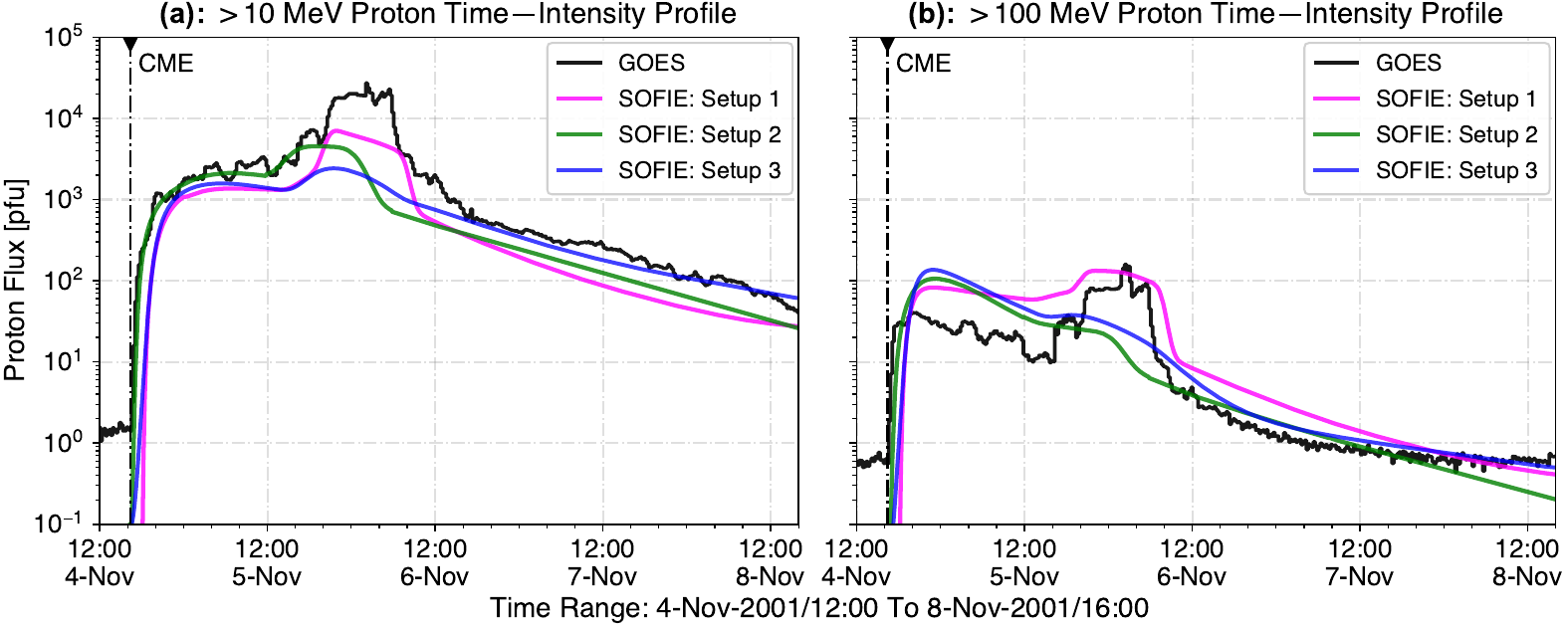}}
    \caption{Comparison of the (a) $>$10 MeV and (b) $>$100 MeV proton time--intensity profiles at Earth observed by GOES (in black) and simulated by SOFIE based on different grid setups (in magenta, green, and blue) in the 4 November 2001 event. A dashed-dotted vertical line marks the launch of the CME in each panel. Detailed explanations of the simulation grid setup are described in text.} \label{fig:compgrid}
    \end{figure*}
  
  In Figure \ref{fig:compgrid}, we compare the $>$10 MeV and $>$100 MeV proton time--intensity profiles between the GOES observations (black) and the SOFIE simulation results based on three grid configurations, Setups 1–3 (magenta, green, and blue, respectively). Across all three runs, both the SEP onset phase during the first day after the eruption and the decay phase after 12:00 UT on 6 November show good agreement with GOES observations. Among them, Setup 1 exhibits the best overall consistency, particularly in reproducing the ESP-phase timing and peak intensity. In contrast, Setup 2, which was run on-site during the SWPT exercise, shows an earlier ESP-phase onset and a smaller peak amplitude. This discrepancy likely arises from the coarser grid resolution in SC, which results in a more diffusive CME evolution. Setup 3, although maintaining finer resolution near the HCS and along the CME- and Earth-directed regions, still produces a slightly reduced ESP-phase peak compared with Setup 1. 
  
\begin{figure*}[ht!]
    \centering{
    \includegraphics[width=0.55\textwidth]{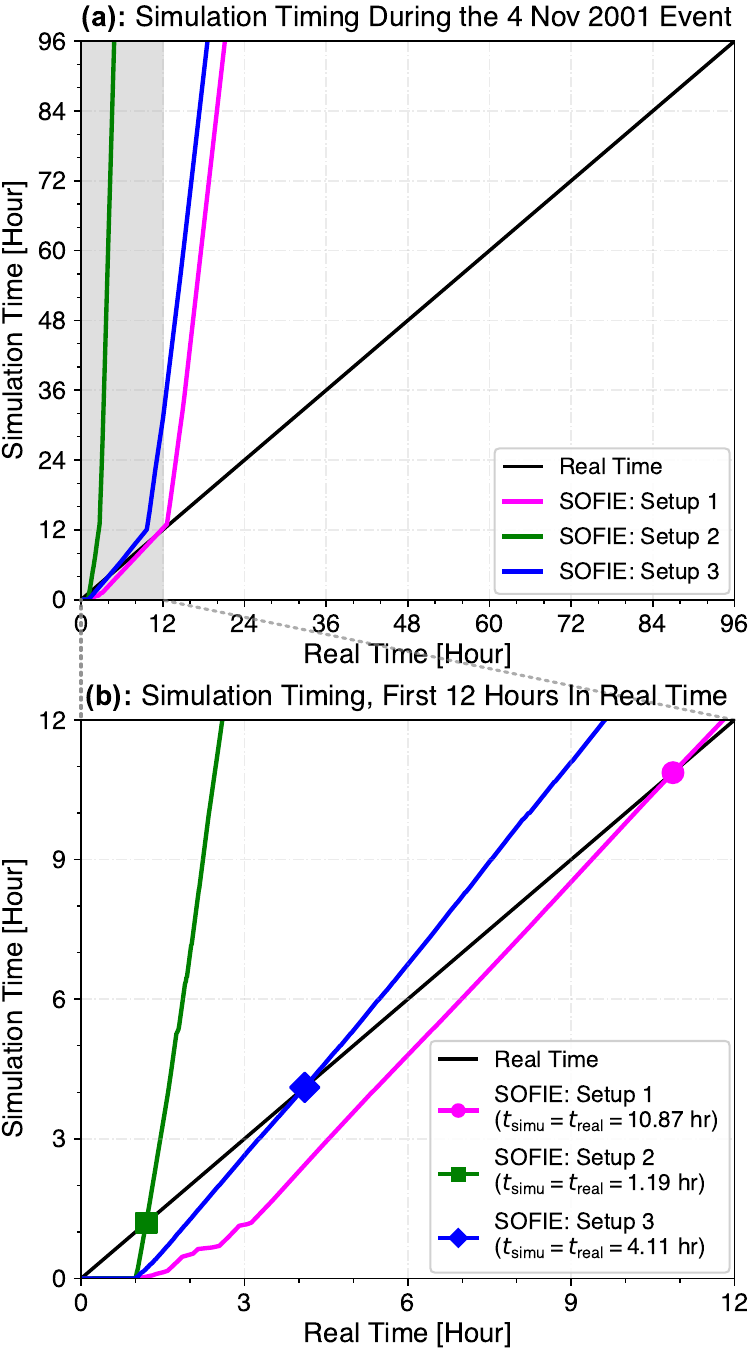}}
    \caption{Simulation timing of SOFIE during the 4 November 2001 SEP event. 
    (a) The overall timing profile over a 4-day window in real time. 
    (b) A zoomed-in view of the first 12 hours in real time. In both panels, the $x$- and $y$-axes represent real time and simulation time, respectively. The magenta, green, and blue curves correspond to SOFIE runtimes based on Setups 1–3, while the black diagonal line represents the real-time reference. In panel (b), the magenta dot, green square, and blue diamond mark the moments when the SOFIE simulation time catches up with real time for the respective setups, with the exact time labeled in the legend. 
    Note that all simulations start after 1 hour of real time, due to the delay in obtaining CME input parameters.} \label{fig:timing}
    \end{figure*}

  In Figure \ref{fig:timing}, we present the simulation runtime of SOFIE using 1,000 CPU cores based on the three grid setups. Panel (a) shows that all runs complete the 4-day simulation faster than in real time, particularly after $\sim$12 hours of simulated time. 
  \textBoldadd{The steepening of the runtime curves after 12 hours of simulated time reflects the changing physical regime throughout the event. Early in the simulation, small grid cells and high fast magnetosonic speeds in the SC domain require small global time steps and frequent SC–IH coupling (30 seconds), as well as short time steps in M-FLAMPA during active shock acceleration. After 12 hours of simulated time, the CME flux rope has largely exited the SC domain and entered the IH domain, the coupling interval is increased to 15 minutes and both MHD and particle time steps increase, substantially reducing the computational cost.} 
  % The computational efficiency of SOFIE in Figure \ref{fig:timing} also reflects the changing physical regime throughout the event. 
  % Early in each simulation, when the flux rope resides in the SC domain, the cell size is small (on the order of $10^{-2}$–$10^{-1}\; R_\mathrm{s}$) and the fast-mode magnetosonic speed is high (hundreds of km s$^{-1}$), requiring small global time steps (on the order of $10^{-1}$ s) in the MHD model. We set a coupling time of 2 minutes between SC and IH, allowing frequent parameter exchanges but also increasing the computational demand. In M-FLAMPA, suprathermal particles are accelerated through the DSA mechansim near the shock front, also requiring small time steps (typically on the order of $10^{-1}$–$10^{0}$ s) that add additional cost. 
  % After 12 hours of the onset, the CME flux rope has fully left the SC domain, so the SC-IH coupling frequency is reduced to 4 hours. The global time step for solving the MHD equations then grows to minutes. In addition, the time step in SP/M-FLAMPA increases to 2 minutes for most field lines, as particle transport dominates over acceleration. Together, all these factors substantially reduce the computational expense at later times. 
  Panel (b) provides a zoomed-in view of the first 12 hours in real time, where the intersections between the SOFIE runtime curves and the real-time reference indicate when the SOFIE simulations catch up with real time, with the $\sim$1-hour delay for obtaining CME input parameters accounted for. We find that Setup 2 reaches this point within 1.19 hours of real time, i.e., 0.19 hours after the simulation starts, while Setups 1 and 3 catch up after 10.87 and 4.11 hours of real time, respectively. 
  \textBoldadd{As shown here, together with Sections \ref{sec3.4:sep2017} and \ref{sec4.4:sep2001}, while the earliest SEP forecast, typically the onset phase, is delayed by a couple of hours due to computational and input latency constraints, SOFIE can still predict the multi-day evolution of SEP fluxes, the decay-phase characteristics, and ESP-phase particle fluxes, providing valuable guidance for the subsequent evolution throughout the SEP event. 
  While we consider two extreme SEP events in the SWPT exercise with sharp particle flux enhancements, the typical time from flare rise or CME onset to SEP threshold crossing ranges from $\sim$1 hour to several hours \cite<e.g.,>[]{papaioannou2016solar, kahler2017forecasting, liu2023AGU, liu2024AGU}; in this situation, the effective latency in the earliest SEP forecasts produced by SOFIE is generally smaller, including for the threshold crossing time, onset peak time, and peak intensity.}
  
    \begin{table*}[htp!]
    \begin{center}
    \caption{Comparison between observations and SOFIE predictions using different grid resolution setups in the 4 November 2001 event.} \label{tab:compres}
    \setlength\tabcolsep{4.3pt}{\hspace*{-0.525cm} % 1.5cm, 1.2ttw
    \adjustbox{width=1.025\textwidth}{
    \begin{tabular}{cc cccc}
        \hline\hline
        \multicolumn{2}{c}{\multirow{2}{*}{Item}} & 
        \multicolumn{4}{c}{4 November 2001 Event} \\ 
        \cline{3-6}
        & & Observation & Setup 1 & Setup 2 & Setup 3 \\ 
        \hline
        \multirow{16}{*}{\shortstack{$>$10 MeV \\\\ Protons}} & Threshold Crossing Time & 2001-11-04T17:05 & 2001-11-04T18:37 & \textbf{2001-11-04T17:11} & 2001-11-04T18:25 \\
        & In Real Time$^*$ & 0.93 hr & 3.92 hr & \textbf{1.13 hr} & 2.53 hr \\
        & Leading Time of SOFIE & -- & $-2.99$ hr & $\boldsymbol{-0.20}$ \textbf{hr} & $-1.60$ hr \\
        \cline{2-6}
        & Onset Peak Time & 2001-11-05T04:45 & 2001-11-05T07:03 & 2001-11-05T06:33 & \textbf{2001-11-05T05:29} \\
        & Onset Peak Intensity & 2,804 pfu & 1,369 pfu & \textbf{2,126 pfu} & 1,579 pfu \\
        & In Real Time$^*$ & 12.60 hr & 12.82 hr & \textbf{2.75 hr} & 9.79 hr \\
        & Leading Time of SOFIE & -- & $-0.22$ hr & $\boldsymbol{+9.85}$ \textbf{hr} & $+2.81$ hr \\
        \cline{2-6}
        & ESP Phase Peak Time & 2001-11-06T02:20 & \textbf{2001-11-05T21:53} & 2001-11-05T18:55 & 2001-11-05T21:31 \\
        & ESP Phase Peak Intensity & 27,149 pfu & \textbf{7,021 pfu} & 4,535 pfu & 2,421 pfu \\
        & In Real Time$^*$ & 34.18 hr & 14.55 hr & \textbf{3.13 hr} & 11.78 hr \\
        & Leading Time of SOFIE & -- & $+19.63$ hr & $\boldsymbol{+31.05}$ \textbf{hr} & $+22.40$ hr \\
        \cline{2-6}
        & Spearman Correlation Coefficient & -- & \textbf{0.9293} & 0.8410 & 0.8681 \\
        & Percentage within an Order of Magnitude & -- & \textbf{99.87\%} & 92.66\% & 84.68\% \\
        & Percentage within a Factor of 2 & -- & \textbf{41.73\%} & 37.71\% & 37.66\% \\
        & Median Logarithmic Error & -- & $-0.1828$ & 0.0542 & $\boldsymbol{-0.0350}$ \\
        & Median Absolute Logarithmic Error & -- & \textbf{0.3323} & 0.4137 & 0.4055 \\
        \hline
        \multirow{16}{*}{\shortstack{$>$100 MeV \\\\ Protons}} & Threshold Crossing Time & 2001-11-04T16:45 & 2001-11-04T18:25 & \textbf{2001-11-04T17:11} & 2001-11-04T18:01 \\
        & In Real Time$^*$ & 0.60 hr & 3.77 hr & \textbf{1.13 hr} & 2.24 hr \\
        & Leading Time of SOFIE & -- & $-3.17$ hr & $\boldsymbol{-0.53}$ \textbf{hr} & $-1.64$ hr \\
        \cline{2-6}
        & Onset Peak Time & 2001-11-04T20:40 & \textbf{2001-11-04T22:55} & 2001-11-04T23:11 & \textbf{2001-11-04T22:55} \\
        & Onset Peak Intensity & 41 pfu & \textbf{82 pfu} & 106 pfu & 136 pfu \\
        & In Real Time$^*$ & 4.52 hr & 7.39 hr & \textbf{1.98 hr} & 5.83 hr \\
        & Leading Time of SOFIE & -- & $-2.87$ hr & $\boldsymbol{+2.54}$ \textbf{hr} & $-1.31$ hr \\
        \cline{2-6}
        & ESP Phase Peak Time & 2001-11-06T02:25 & \textbf{2001-11-05T22:09} & 2001-11-05T16:27 & 2001-11-05T18:41 \\
        & ESP Phase Peak Intensity & 162 pfu & \textbf{132 pfu} & 28 pfu & 38 pfu \\
        & In Real Time$^*$ & 34.27 hr & 14.58 hr & \textbf{3.06 hr} & 11.42 hr \\
        & Leading Time of SOFIE & -- & $+19.69$ hr & $\boldsymbol{+31.21}$ \textbf{hr} & $+22.85$ hr \\
        \cline{2-6}
        & Spearman Correlation Coefficient & -- & \textbf{0.9178} & 0.5628 & 0.5716 \\
        & Percentage within an Order of Magnitude & -- & \textbf{99.63\%} & 92.09\% & 97.81\% \\
        & Percentage within a Factor of 2 & -- & 41.69\% & \textbf{49.22\%} & 46.74\% \\
        & Median Logarithmic Error & -- & 0.3431 & \textbf{0.1299} & 0.2136 \\
        & Median Absolute Logarithmic Error & -- & 0.3563 & \textbf{0.3029} & 0.3615 \\
        \hline
        \multicolumn{2}{c}{Real Time to Provide 4-day Proton Fluxes} & 96.00 hr & 21.12 hr & \textbf{4.86 hr} & 18.57 hr \\
        \multicolumn{2}{c}{Leading Time of SOFIE} & -- & $+74.88$ hr & $\boldsymbol{+91.14}$ \textbf{hr} & $+77.43$ hr \\
        \hline
        \multicolumn{2}{c}{When can SOFIE catch up with real time?} & -- & 10.87 hr & \textbf{1.19} \textbf{hr} & 4.11 hr \\
        \hline
        \multicolumn{6}{l}{\textit{Note.} The three simulation setups listed in the header correspond to different grid resolution configurations, as described in text. Items } \\
        \multicolumn{6}{l}{marked with $^*$ indicate the real time when each result was obtained, representing the elapsed time since the flare onset (2001-11-04T16:09). } \\
        \multicolumn{6}{l}{The prediction leading time is calculated as the difference between the real time at which the simulation result was obtained and the } \\
        \multicolumn{6}{l}{corresponding observation time. A positive value indicates that SOFIE ran faster than real time, and vice versa. For each grid setup and } \\
        \multicolumn{6}{l}{energy channel, we use multiple metrics to evaluate the overall time--intensity profile, and include the moment when the SOFIE simulation } \\
        \multicolumn{6}{l}{time caught up with the real time. Besides, for each item, the optimal result or shortest timing among the three setups is highlighted in bold. } \\
    \end{tabular}}}
    \end{center}
    \end{table*}
  
  Moreover, based on the simulation results and runtimes \textBoldadd{for the 4 November 2001 event}, Table \ref{tab:compres} details the threshold crossing time, the onset peak time and intensity, and the ESP phase peak time and intensity in the $>$10 MeV and $>$100 MeV proton fluxes as measured by GOES and simulated by SOFIE under Setups 1–3. For both energy channels, five statistical metrics commonly used in SEP model validation \cite<e.g.,>[and \url{https://ccmc.gsfc.nasa.gov/challenges/sep/validation/}]{whitman2023review, whitman2024multi, whitman2026validation} are also calculated to evaluate the 4-day time profiles. Additionally, we list the real time when GOES observed or SOFIE predicted each item. In the following, we summarize the key findings from the comparison \textBoldadd{for the 4 November 2001 event}. 
  \begin{itemize}
    \item[1.] Overall, Setup 1 provides the best agreement with GOES observations, particularly in reproducing the ESP phase, although it requires $\sim$21 hours to complete the 4-day prediction. Specifically, it achieves a point-to-point Spearman correlation coefficient $>$0.9, with $>$99\% of data points within an order of magnitude and about 42\% within a factor of 2 of the GOES measurements. 
    \item[2.] Setup 2 runs substantially faster than real time ($\sim$5 hours), while achieving comparable accuracy and capturing the main temporal and intensity trends. \textBoldadd{Compared with the other two setups, Setup 2 exhibits a noticeably reduced delay in reproducing the early SEP threshold crossings and reaches the modeled onset peak several hours earlier in real time.} $\sim$92\% of its data points fall within an order of magnitude and $\sim$40\% (38\% for $>$10 MeV and 49\% for $>$100 MeV protons) within a factor of 2 of the GOES observed data. 
    \item[3.] Setup 3 performs moderately between the two in terms of runtime\textBoldadd{. For the 4 November 2001 event presented here}, \textBoldmod{and its results can occasionally be}{results based on Setup 3 are slightly} better \textBoldadd{in some metrics}, but are typically intermediate or show larger deviations from the GOES measurements than Setups 1 and 2. Its minimal global time step is comparable to that of Setup 1, explaining the similar completion time for the 4-day simulation. However, its coarser background grid resolution, as used in Setup 2, reduces the accuracy of CME and SEP predictions. 
    \item[4.] For all setups, the $>$10 MeV predictions exhibit a bias (median logarithmic error) close to zero, suggesting that the predicted values are generally centered around the GOES observations. In contrast, the $>$100 MeV predictions show a positive bias, primarily due to the overestimation of the onset peak as shown in Figure \ref{fig:compgrid}(b). The aggregated accuracy (median absolute logarithmic error) is similar among all setups, indicating that half of the predicted data points fall within about one-third of an order of magnitude of GOES measurements. 
  \end{itemize}
  These results demonstrate that all three setups are capable of providing operationally useful forecasts, with Setup 1 offering higher accuracy and Setup 2 providing timely, \textBoldmod{decision-referencing}{operationally referencing} predictions that can operate significantly faster than real time. Based on the comparison above, for human space missions like Artemis II, it is recommended to run Setups 1 and 2 simultaneously: Setup 2 provides a good order-of-magnitude estimate of the SEP event within a couple of hours, while Setup 1 increases the accuracy. Setup 2 should be used in the first $\sim$5 hours or so, and Setup 1 will take over afterwards. Nevertheless, further validation studies and event-by-event analysis should be conducted to provide generic suggestions on the grid resolution configuration used for operations.

  \subsection{Uncertainties in Predicted Results} \label{sec5.2:uncertainty}
  
  Although SOFIE has been applied to multiple historical SEP events \cite{zhao2024solar} and tested under simulated \textBoldmod{real-time}{operational} conditions during the SWPT exercise, some deviations from observations remain. For the three components within SOFIE, we list their input free parameters used in this work in Table \ref{tab:param}, respectively. %with the Poynting flux parameter set following \citeA{huang2024solar} and the CME flux rope parameters derived from the AR location and CME speed. 
  Uncertainties of the predicted SEP fluxes may arise from these input parameters, such as the background solar wind based on parameter tuning for historical events and the CME input parameters. Although not explored here, assessing the sensitivity of SOFIE's SEP predictions to variations in CME parameters, such as those derived from different sources (e.g., SWPC, M2M SWAO, or LASCO catalog), would be valuable for quantifying model robustness and guiding its future operational implementation \textBoldadd{\cite<see also>[]{mays2015ensemble, richardson2015properties, papaioannou2025exploring}}. 
  In the following, we focus our discussion on the free input parameters used in M-FLAMPA and their influence on the accuracy and reliability of the predicted results. % outline potential improvements for future scientific studies and operational applications. 
  
  In M-FLAMPA, seed particles are estimated following the suprathermal tail of the solar wind and injected uniformly at the shock front using a scaling factor. This scaling factor is a free parameter set as 1.0 by default, which can be highly affected by the properties of the CME-driven shock \cite<see Section 4.5 in>[for its physical interpretations]{zhao2024solar}. In practice, it may need to be fine-tuned to reproduce measured SEP fluxes. For example, we adopted a scaling factor of 10 for the 4 November 2001 event in the SWPT exercise to better reproduce the GOES measurements. 
  \textBoldadd{Because this parameter primarily controls the overall particle intensity level, its effective value is typically constrained when the SEP onset peak is issued. Together with the delayed availability of the first forecast in the initial hours, this represents a current operational limitation. Nevertheless, once constrained, the model provides valuable information on the subsequent evolution of SEP intensities, decay-phase characteristics, and particle fluxes during the ESP phase.} For operational predictions, however, a more robust estimate of seed particle populations injected into the particle acceleration process \cite<e.g.,>[]{zank2000particle, guo2013acceleration, caprioli2014simulations, afanasiev2015self} is needed and essential to improve the predictive capability of SEPs \textBoldadd{in the future}. 
  
  Another free parameter in M-FLAMPA is the \textBoldadd{parallel} mean free path parameter of SEPs in the upstream solar wind, which affects both particle acceleration at the shock and subsequent transport processes. 
  Variations in this parameter can alter the predicted threshold crossing time, peak intensity, decay phase profile, and particle energy spectrum (see Figure 9 in \citeA{zhao2024solar}, Figure 15 in \citeA{liu2025physics}, \textBoldadd{\citeA{els2024diffusion, lang2024detailed, zhong2024mean}, and references therein}). 
  In the SWPT exercise, we used the same value of 0.1 au for both events (see Table \ref{tab:param}), although the optimal value may differ for individual events. 
  In the future, we plan to simulate \textBoldmod{more}{a larger set of} historical events during the SOFIE validation phase, allowing this free parameter to be adjusted based on historical event statistics or empirical relations to improve forecast reliability for operational predictions. 
  
  \textBoldadd{In the current configuration of M-FLAMPA, we also note that cross-field diffusion and particle drift effects are neglected (see Section \ref{sec2:method} and references therein), such that SEP transport is restricted to individual magnetic field lines. As a result, uncertainties in magnetic connectivity (e.g., the event shown in Figure \ref{fig:cmeshk201709}(e)) can further influence the predicted SEP fluxes. 
  Under these assumptions, the parallel mean free path mentioned here should be regarded as an effective parameter rather than a direct representation of the physical scattering conditions when additional transport processes are included. While addressing these limitations will require further model development, ensemble simulations and probabilistic estimates of the mean free path represent promising refinements to improve forecast robustness and reliability for current operational applications.}
  % While SOFIE captures the overall trends of SEP events in this and previous studies, these uncertainties underscore the need for continued model validation to improve its predictive performance. 

\section{Summary and Conclusion} \label{sec6:sumcon}

As SEPs can pose significant space radiation hazards, it is crucial to provide a timely prediction of their fluxes at any location in interplanetary space for upcoming human exploration missions. The physics-based models can play an important role for this purpose. In this work, we describe the simulated \textBoldmod{real-time}{operational} testing of SOFIE, a physics-based SEP model developed in the CLEAR Space Weather Center of Excellence, and demonstrate its performance in the 2025 SWPT exercise at NOAA/SWPC. 

Designed to predict SEP fluxes, SOFIE self-consistently integrates models that simulate the ambient solar wind, parameterize CME flux ropes, and solve shock-driven particle acceleration and transport processes. 
During the SWPT exercise, SOFIE was tested under simulated \textBoldmod{real-time}{operational} conditions for two historical SEP events, the 10 September 2017 and 4 November 2001 eruptions. 
For the 10 September 2017 event, we used the default grid configuration previously employed in \citeA{zhao2024solar} and \citeA{liu2025physics}, where SOFIE could provide accurate 4-day SEP predictions within 13 hours of real time. 
For the 4 November 2001 event, based on in-depth discussions with analysts and operators from SWPC, CCMC, SRAG, and M2M SWAO, we coarsened the grid in \textBoldadd{the} SC by a factor of 2 and performed on-site simulations at the SWPT. Running on 1,000 CPU cores, SOFIE reproduced key features of the observed CME coronagraph images and SEP fluxes, and was able to deliver 4-day predictions in only 5 hours, which is significantly faster than real time. 
Note that for both events, we prepared the background solar wind in advance, and the corresponding timing is not included. For \textBoldmod{operations}{operational use}, the steady-state solar wind has been simulated every day through an established pipeline \cite<\url{https://solarwind.engin.umich.edu/current/},>[]{sokolov2025_pipeline} driven by the most up-to-date GONG photospheric \textBoldmod{magnetogram}{synoptic map}. When a CME is detected, we can restart the simulation from the \textBoldmod{saved}{latest available} background solar wind. \textBoldadd{While we acknowledge that the solar wind can evolve within a 24-hour window and that preceding CMEs may modify the heliospheric background and magnetic connectivity \cite<e.g.,>[]{manchester2017physical, zhuang2020role}, a detailed assessment of these effects is beyond the scope of the present study and will be addressed in future work, including the use of a shorter and computationally affordable cadence of background update.} 
Moreover, interactive discussions on-site at the SWPT also suggested coarsening only the background grid in \textBoldadd{the} SC \textBoldadd{domain} while retaining the finer resolution of the HCS, CME path, and Earth-directed regions as the default setup. 
Therefore, we conducted post-exercise simulations and comparisons among three grid setups, showing that the default grid achieves the highest accuracy, while the coarser grid adopted on-site during the SWPT exercise offers timely, decision-oriented predictions in applications.

% Compared with the grid setup for scientific studies, the AMR criteria in the SC domain was restricted to the HCS, Earth cone, and along the CME path, while other regions were coarsened in order to improve runtime efficiency while preserving accuracy, based on in-depth discussions with analysts and operators from SWPC, CCMC, SRAG and M2M office. 
% Running on 1,000 CPU cores of the NASA Pleiades supercomputer, SOFIE reproduced key features of the observed CME coronagraph images and SEP fluxes within a short timing, and was able to provide 4-day prediction results significantly faster than real time.

These results for the SWPT exercise demonstrate that the physics-based SEP model, SOFIE, traditionally viewed as computationally expensive, can in fact meet the latency and robustness requirements of operational predictions. With continued validation and integration into model forecasting, SOFIE represents a significant step toward the real-time, physics-based SEP prediction in support of future human space exploration.

%%%%%%%%%%%%%%%%%%%%%%%%%%%%%%%%%%%%%%%%%%%%%%%
% DATA SECTION and ACKNOWLEDGMENTS
%%%%%%%%%%%%%%%%%%%%%%%%%%%%%%%%%%%%%%%%%%%%%%%

\section*{Open Research Section}

% Data Availability Statement
The GONG and MDI \textBoldmod{magnetograms}{synoptic maps} used in this work are available on \url{https://gong.nso.edu/data/magmap/} and \url{http://jsoc.stanford.edu/MDI/MDI_Magnetograms.html}, respectively. 
The SOHO/LASCO WL image data are obtained using SolarSoftWare (SSW, \url{https://www.lmsal.com/solarsoft/}), which downloads original data through the Virtual Solar Observatory (VSO, \url{https://sdac.virtualsolar.org/cgi/search}) and accounts for the instrumental calibration. 
The \textit{in situ} ACE solar wind plasma and magnetic field parameters used in this study are accessible on \url{https://izw1.caltech.edu/ACE/ASC/level2/index.html}. 
The GOES SEP data are obtained from \url{https://www.ncei.noaa.gov/data/goes-space-environment-monitor/access/avg/}. 
The SWMF/SOFIE is available on \url{https://github.com/SWMFsoftware}, and the SOFIE model can be run via run-on-request (\url{https://ccmc.gsfc.nasa.gov/ror/requests/SH/SWMF-AWSoM/swmfawsom_user_registration.php}) at CCMC. 
All observation and simulation data shown in plots, including the 2D ecliptic plane data, time series of solar wind plasma parameters along Earth's trajectory, 3D solar wind plasma and magnetic field data right after flux rope insertion, WL image data, and 2D distributions and time--intensity profiles of energetic protons, are publicly available on the Zenodo platform \cite{liu2025tbe_zenodo}.

\section*{Conflict of Interest}

The authors declare that there are no conflicts of interest for this study.

\acknowledgments
\textBoldadd{The authors sincerely thank the two anonymous reviewers for their time and effort devoted to improving this paper.} 
This work is supported by NASA Living With a Star (LWS) Strategic Capability project under NASA grant 80NSSC22K0892 (SCEPTER), NASA Space Weather Center of Excellence program under award 80NSSC23M0191 (CLEAR), NSF grant GEO-2149771 (ANSWERS), NASA grant 80NSSC21K1124, and NASA LWS grant 80NSSC20K1778. 
% K. M. acknowledges support from NOAA grant NA22OAR4320151 for the Cooperative Institute for Earth System Research and Data Science. 
The authors express their great gratitude to the SOHO project, an international cooperation between \textBoldmod{ESA}{European Space Agency (ESA)} and NASA, the GOES and ACE teams, and also the NSO Integrated Synoptic Program. 
The authors also thank Dr. Yang Liu from Stanford University for providing assistance with the MDI magnetogram data. 
Computational resources supporting this work are provided by the NASA High-End Computing (HEC) Program through the NASA Advanced Supercomputing (NAS) Division at the Ames Research Center (\url{https://www.nas.nasa.gov/hecc/}), and the high-performance computing support from the Texas Advanced Computing Center (TACC) Frontera at the University of Texas at Austin \cite<\url{https://tacc.utexas.edu/},>[]{stanzione2020frontera}. 
Any opinions, findings, conclusions or recommendations expressed in this material are those of the authors and do not necessarily reflect the views of the participating agencies or affiliated institutions.

%%%%%%%%%%%%%%%%%%%%%%%%%%%%%%%%%%%%%%%%%%%%%%%
% REFERENCES and BIBLIOGRAPHY
% \bibliography{<name of your .bib file>} don't specify the file extension
% don't specify bibliographystyle
%%%%%%%%%%%%%%%%%%%%%%%%%%%%%%%%%%%%%%%%%%%%%%%

\bibliography{ref_paper}

\end{document}